\documentclass[aps,prd,showpacs,twocolumn,nofootinbib,superscriptaddress]{revtex4}
\usepackage[usenames]{color}
\usepackage{graphicx}
\usepackage{amssymb}
\usepackage{hyperref}

\usepackage{amsfonts}
\usepackage{amsmath}
\usepackage{amsthm}
\usepackage{bm}
\usepackage{dcolumn}
\usepackage{epsfig}
\usepackage{graphicx}
\usepackage{graphics}
\usepackage[latin1]{inputenc}
\usepackage{latexsym}
\usepackage{rotating}

\newcommand\calF{{\cal F}}

\newcommand\calAmp{{\cal G}}
\newcommand\calAmpMatch{\tilde{{\cal G}}}
\newcommand\unit[1]{{\rm #1}}

\newcommand\optional[1]{}
\newcommand{\lamRL}{\lambda_{\mbox{\tiny R,L}}}
\newcommand{\RL}{{\mbox{\tiny R,L}}}
\newcommand{\R}{{\mbox{\tiny R}}}
\newcommand{\LL}{{\mbox{\tiny L}}}
\newcommand{\FRW}{{\mbox{\tiny FRW}}}
\newcommand{\GR}{{\mbox{\tiny GR}}}
\newcommand{\FLAT}{{\mbox{\tiny FLAT}}}

\begin{document}

\title{Testing gravitational parity violation with coincident \\ gravitational waves and short gamma-ray
bursts}

\author{Nicol\'as Yunes}
\affiliation{Department of Physics, Princeton University, Princeton, NJ 08544, USA.}
\affiliation{Department of Physics, Institute for Gravitation and the Cosmos, 
The Pennsylvania State University, University Park, PA 16802, USA.}

\author{Richard O'Shaughnessy}
\affiliation{Department of Physics, Institute for Gravitation and the Cosmos, 
The Pennsylvania State University, University Park, PA 16802, USA.}

\author{Benjamin J. Owen}
\affiliation{Department of Physics, Institute for Gravitation and the Cosmos, 
The Pennsylvania State University, University Park, PA 16802, USA.}
\affiliation{Max Planck Institut f\"ur Gravitationsphysik (Albert Einstein
Institut), Callinstr.\ 38, 30167 Hannover, Germany}

\author{Stephon Alexander}
\affiliation{Department of Physics, Institute for Gravitation and the Cosmos, 
The Pennsylvania State University, University Park, PA 16802, USA.}
\affiliation{Department of Physics and Astronomy, Haverford College, Haverford, PA 19041, USA.}

\begin{abstract}

Gravitational parity violation is a possibility motivated by
particle physics, string theory and loop quantum gravity.
One effect of it is amplitude birefringence of gravitational waves, whereby
left and right circularly-polarized waves propagate at the same speed
but with different amplitude evolution.
Here we propose a test of this effect through coincident observations of
gravitational waves and short gamma-ray bursts from binary mergers involving
neutron stars.
Such gravitational waves are highly left or right circularly-polarized
due to the geometry of the merger.
Using localization information from the gamma-ray burst,  
ground-based gravitational wave detectors can measure the distance to the source with reasonable accuracy.
An electromagnetic determination of the redshift from an afterglow or host
galaxy yields an independent measure of this distance.
Gravitational parity violation would manifest itself as a discrepancy between 
these two distance measurements.
We exemplify such a test by considering one specific effective theory that 
leads to such gravitational parity-violation, Chern-Simons gravity.  
We show that the advanced LIGO-Virgo network and all-sky gamma-ray telescopes 
can be sensitive to the propagating sector of Chern-Simons gravitational parity violation
to a level roughly two orders of magnitude better than current stationary constraints from the 
LAGEOS satellites.

\end{abstract}


\maketitle

\section{Introduction}

Several reasons exist to believe that there is some degree of parity
violation in gravity, at least at the quantum scale and possibly carrying
through to macroscopic scales through the anomaly mechanism.
From an experimental standpoint, it is curious that the weak interaction
exhibits maximal parity violation, yet the other fundamental forces
apparently exhibit none. From a theoretical standpoint, parity violation arises from the standard
model of particle physics~\cite{Bell:1969ts,1969PhRv..177.2426A,AlvarezGaume:1983ig}, 
some sectors of string theory~\cite{Green:1987mn,Alexander:2004xd}, 
and extensions of loop quantum gravity~\cite{Ashtekar:1988sw,Taveras:2008yf,Calcagni:2009xz,Mercuri:2009zt,Gates:2009pt}.

Gravitational parity violation can be encoded in the action as an extension to general relativity (GR)
that consists of the addition of an antisymmetric product of curvature tensors coupled to a scalar field, 
which to leading order in the curvature leads uniquely to Chern-Simons (CS) gravity~\cite{Alexander:2009tp}.
Jackiw and Pi~\cite{Jackiw:2003pm} demonstrated that the CS term can be
covariantly embedded with its three dimensional counterpart as a consistent
modification to GR.
In the standard model, there exists a radiatively generated one-loop chiral 
anomaly that gives rise to a CS term coupled to lepton number~\cite{Weinberg:1996kr}.
In heterotic and Type I superstring theories, the CS term is generated through the 
Green-Schwarz gauge anomaly-canceling mechanism~\cite{Polchinski:1998rr}. 
In loop quantum gravity, the scalarization of the Barbero-Immirzi parameter coupled
to fermions has been shown to lead to an effective CS 
action~\cite{Ashtekar:1988sw,Taveras:2008yf,Calcagni:2009xz,Mercuri:2009zt,Gates:2009pt}.
Even devoid of a specific fundamental theory, one can also show that the CS term
unavoidably arises in effective field theories of inflation as one of the few
non-vanishing, second-order curvature corrections to the
Einstein-Hilbert action~\cite{Weinberg:2008hq}.
Another phenomenological parity-violating extension to GR has been
proposed where the fundamental constants of nature (e.g.,~Newton's constant)
break parity invariance~\cite{Contaldi:2008yz}.

Most investigations of gravitational wave (GW) signatures of parity violation
have focused on {\emph{amplitude birefringence}}.
That is, compared to GR, right circularly-polarized waves are enhanced or suppressed and left circularly-polarized
waves are suppressed or enhanced as they propagate~\cite{Jackiw:2003pm,Lue:1998mq,Alexander:2007kv,Yunes:2008bu,Alexander:2009tp}.
Such birefringence occurs when GWs propagate on a flat or curved background. 
In fact, the propagation of such waves over cosmological distances has been proposed as an explanation 
of the baryogenesis problem in the early Universe~\cite{Alexander:2004us} and could have observable effects on
the cosmic microwave background~\cite{Alexander:2004wk, Saito:2007kt,
Contaldi:2008yz} and a stochastic GW background~\cite{Seto:2007tn,
Satoh:2007gn}. Cosmological GW amplitude birefringence has been discussed 
in the context of GW observations of cosmological supermassive binary black holes (BHs) with the
planned Laser Interferometer Space Antenna (LISA)~\cite{Alexander:2007kv}.

Amplitude birefringence arises in propagating modes if the parity operator does
not commute with the Hamiltonian. 
In its most general form, the effect of the plane-wave propagator can be expressed as
\begin{equation}
\label{matrix}
\left( \begin{array}{c} {h_{+,k}(t)} \\ h_{\times,k}(t) \end{array} \right)
= e^{-i\omega t}
\left( \begin{array}{rr} u & iv \\ -iv & u \end{array} \right)
\left( \begin{array}{c} h_{+,k}(0) \\ h_{\times,k}(0) \end{array} \right).
\end{equation}
Here $\omega$ is the GW angular frequency, $t$ is some time coordinate,
$h_{+,\times,k}$ are the Fourier components of the GW of wavenumber $k$, and $u$ and $v$
are multiplicative factors characterizing the translation invariant amplification/suppression 
of each Fourier mode during propagation in  time. 
The quantity $u$ corrects for background curvature effects: $u = 1$ in a flat-background, but it acquires redshift corrections in a
cosmological background (see e.g.~\cite{Laguna:2009re}). 
Given a specific modified theory, Eq.~\eqref{matrix} can be obtained by solving
the modified field equations for the propagating modes of the metric perturbation, where 
$v$ measures the degree of parity violation.
The right- and left-circular polarizations $h_{\RL} = (h_+ \pm ih_\times) /
\sqrt{2}$ then propagate as
\begin{equation}
\label{matrix2}
\left( \begin{array}{c} h_{\R,k}(t) \\ h_{\LL,k}(t) \end{array} \right)
= e^{-i\omega t}
\left( \begin{array}{rr} u + v & 0 \\ 0 & u-v \end{array} \right)
\left( \begin{array}{c} h_{\R,k}(0) \\ h_{\LL,k}(0) \end{array} \right).
\end{equation}
In the presence of amplitude birefringence, the eigenvalue matrix of the propagator operator 
remains diagonal, but with modified eigenvalues. Clearly then,
right circularly-polarized waves are amplified or suppressed, while left circularly-polarized 
waves are suppressed or enhanced relative to GR, depending on the sign of $v$.

The amplitude birefringence discussed above and in the rest of this paper is a pure propagation effect, which accumulates
over propagation distance, in contrast with  wave generation effects which do not.  As the latter cause parity
violation ``at the source,'' their effect is observable at 
 any 
distance~\cite{Jackiw:2003pm,Yunes:2007ss}.  Wave generation modifications also couple to matter as well as
curvature~\cite{Alexander:2008wi}, for example leading to weakly composition dependent waveforms   in BH-NS versus
NS-NS mergers.
Recently,~\cite{Sopuerta:2009iy} studied parity violation in wave generation for extreme-mass ratio inspirals, finding that indeed trajectories 
and the multipolar generation scheme itself encode such violation at a fundamental level.
The influence of such wave generation effects on GW observables would require a full (non-extreme mass ratio)
post-Newtonian and multipolar analysis, which is beyond the scope of this paper.  Instead, here we concentrate on 
wave propagation effects only and for simplicity adopt pure GR wave
generation without modified-gravity corrections, recognizing that 
(i) such wave generation effects are calculable on a theory-by-theory basis and 
(ii) their distance-independent influence on binary phase should be easily distinguishable from any secular amplification accumulated
over cosmological distances.\optional{\footnote{For example, depending on the theory waveform generation effects will
    modify the binary orbital phase and thus duration
    comparable to some post-Newtonian order.   On the one hand, modified waveform duration impacts overall signal
    fourier amplitude (through the
    stationary-phase fourier transform $\tilde{h}\propto (df/dt)^{-1/2}$).  On the other hand, gravitational wave data
    analysis strategies are highly sensitive to orbital phase.  Any orbital phase modification comparable  can be easily identified
    and extracted, to a precision comparable to the relevant post-Newtonian order.  
We anticipate a detailed analysis with waveform generation effects will produce a Fisher matrix that allows measurements
roughly as described in the text; see Footnote \ref{foot:Fisher:WhatWeShouldDo} for further discussion.
}}

In this paper, we show that an interesting test of gravitational parity violation due to propagation
effects can be made with existing instruments: the SWIFT~\cite{Gehrels:2004am} and 
GLAST/Fermi~\cite{Carson:2006af} gamma-ray satellites, and ground-based LIGO~\cite{Abbott:2007kv} and
Virgo~\cite{Acernese:2007zze} GW detectors after ongoing upgrades.
Such a test relies on the detection of coincident GW/gamma-ray burst (GRB) events, a possible progenitor
of which are neutron star (NS) binary mergers, that lead to a redshift observation either from afterglow or host 
galaxy identification.
The reasons why such events are excellent to constrain parity violation are the same as those outlined in 
work that proposes the use of GW/GRB events to constrain GRB parameters~\cite{Seto:2005xb} and cosmological
parameters~\cite{Dalal:2006qt,Nissanke:2009kt,Laguna:2009re}.
Due to the collimation of the jet producing the GRB, the binary's orbital
angular momentum at merger must point along the line of sight to Earth, and
thus the GW signal must be highly circularly-polarized.
Together with the accurate sky location obtained from the GRB, this allows
an accurate distance measurement from the GW signal. This distance measurement can then 
be compared to the purely electromagnetic distance measurement,
which for typical redshifts $z \sim 0.1$ depends only on the Hubble constant.
Amplitude birefringence will manifest as a discrepancy between these two
distance measurements.
If no discrepancy is found, the error on the distance measurements can be used to place
upper limits on the possible degree of gravitational parity violation.

The test proposed here is unique in that it employs only
the radiative sector of any generic, gravitational parity violating theory. Current constraints
on gravitational parity violation concentrate on a specific alternative theory (CS gravity) 
and explore only its stationary sector through Solar System experiments~\cite{Alexander:2007zg, Alexander:2007vt, Smith:2007jm} 
and binary pulsar tests~\cite{Yunes:2008ua}. As such, these tests constrain only specific and local deviations from gravitational parity invariance
in the neighborhood of the gravitational source -- that is, near Earth in the Solar System case and near J0737-3039 in
the binary pulsar case. The test we discuss here constrains {\emph{generic parity violation}} over distances of hundreds of Mpc, 
along the light-cone on which the amplitude birefringent GWs propagate.  

This test also compares favorably to future GW tests with LISA~\cite{Alexander:2007kv, Yunes:2008bu}.
The fractional error in a distance measurement is inversely proportional to the signal-to-noise ratio (SNR) $\rho$.
Although the latter is larger in the LISA case, the sensitivity to the parity violating effect is greater here, because this effect
also depends on the GW frequency, which is much larger during NS mergers.
Propagation effects should accumulate with distance propagated, and on
dimensional grounds this distance should be measured in gravitational
wavelengths -- that is, the effect should be proportional to $D \, f$, where
$D$ is the distance propagated and $f$ is the GW frequency. 
While GW/GRB events will have smaller $D$s and lower $\rho$s than LISA
supermassive BH observations (both by 1--2 orders of magnitude), 
$f$ is 5--6 orders of magnitude higher for the LIGO-Virgo network.
Thus, the sensitivity to parity violation could be 2--3 orders of magnitude
better than in the LISA case, although as we shall see later this is somewhat reduced by
systematic errors.

The straw-man chosen here to compare against existing tests of gravitational parity violation 
is CS gravity. In this theory, the magnitude of gravitational parity violation 
is controlled by a length parameter (in geometric units), proportional to certain time derivatives of the 
CS coupling field $\theta$. Solar system observations of the LAGEOS satellites~\cite{Smith:2007jm} 
have placed the bound $\dot{\theta}_{0} < 2000 \; {\rm{km}}$ at $2\sigma$ when
$\ddot{\theta}_{0} = 0$. (The subscripts refer to evaluation at present, i.e.\ zero redshift.)
We have found that the above constraint can be generalized assuming $\theta$ evolves on cosmological
timescales. Allowing for both $(\dot{\theta}_{0}, \ddot{\theta}_{0})$ to be nonzero to leading order,  
we can generalize the Solar System constraint via the replacement rule
$\dot{\theta}_{0} \to |\dot{\theta}_{0} - \ddot{\theta}_{0}/H_{0}|$, which then
leads to the much more powerful bound  
$|\dot{\theta}_{0} - \ddot{\theta}_{0}/H_{0}| < 2000 \; {\rm{km}}$.

\begin{figure}
\includegraphics[width=\columnwidth,clip=true]{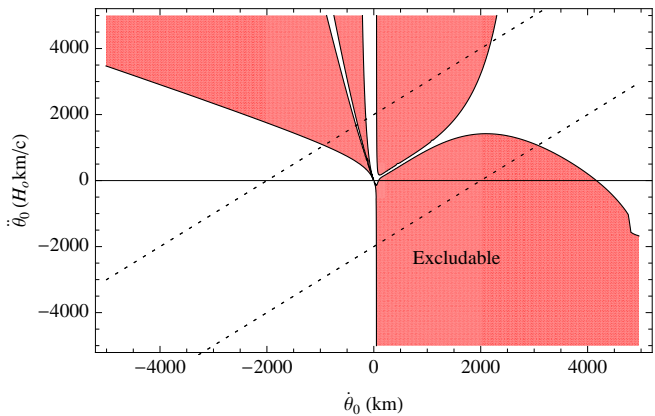}
\caption{\label{fig:intro} Given a coincident right-handed GW/GRB detection at an SNR of $10$, 
the gray-shaded region (red in the color version) shows the area of $(\dot{\theta}_{0},\ddot{\theta}_{0})$
parameter space that could be excluded at $2\sigma$ by a measurement of one right-handed
short GRB merger waveform consistent with its expected GR amplitude, given host galaxy distance 
and (measured) chirp mass.  The region excluded by Solar System observations
($|\dot{\theta}_{0} - \ddot{\theta}_{0}/H_{0}| < 2000 \; {\rm{km}}$ at $2\sigma$) is outside the dotted lines.}
\end{figure}

Our calculations suggest that the LIGO-Virgo network will be dramatically more
sensitive to parity violation, as we show in Fig.~\ref{fig:intro} for a
right-handed GW observation. The gray-shaded regions 
(red in the color version) show the areas of phase space that could be detected or
excluded with a coincident LIGO GW/GRB observation at $2\sigma$. This is to be compared to the
regions of phase space ruled out by the generalized Solar System constraints, denoted in the figure
as the region outside dotted lines. Inside the dotted lines, the generalized Solar System constraint
cannot rule out the CS modification. 
A GW observation could place much tighter constraints on the allowed region of parity violating phase space. 
Such constraints are interesting because gravitational parity violation need not
be Planck suppressed in the fundamental theory from which CS gravity 
derives. This is because the effective CS theory is shift-invariant, and thus, the scalar field 
that sources the correction must be massless, if this symmetry is not to be broken. 

A certain asymmetry in the possible constraints we could place on $(\dot{\theta}_{0},\ddot{\theta}_{0})$ 
given a single, coincident GW/GRB detection is clearly discernible in Fig.~\ref{fig:intro}. 
This asymmetry is an artifact of assuming a purely {\emph{right-polarized}} GW event. 
In CS gravity, right-polarized signals possess a resonant behavior if $\dot{\theta}_{0} > 0$ 
for frequencies in the LIGO sensitivity band, which greatly enhances the amount of parity 
violation and the constraints one could place. This situation, however, reverses if instead we 
had assumed a left-polarized GW observation. In such a case, we would have obtained exclusion 
regions identical to those of Fig.~\ref{fig:intro} but reflected under 
$(\dot{\theta}_{0},\ddot{\theta}_{0}) \to (-\dot{\theta}_{0},-\ddot{\theta}_{0})$.
Therefore, a two-detector observation of a right and a left-polarized GW would allow us to 
constrain $(\dot{\theta}_{0},\ddot\theta_{0})$ equally well for both positive and negative 
$\dot{\theta}_{0}$, thus extending the exclusion regions of Fig.~\ref{fig:intro}.

All the results we just described assume a somewhat idealistic test, where we have ignored certain astrophysical uncertainties, 
such as inaccuracies in host galaxy identification. Perhaps one of the most important uncertainties relates to the inclination angle.
GRB jets are not necessarily perfectly aligned with the line of sight, which implies that the inclination angle is not exactly known 
{\emph{a priori}} and the GW signal is not perfectly circularly polarized. Since the inclination angle is degenerate with the luminosity 
distance in the GW response function, this deteriorates GW measurements of the latter by roughly a factor of two. 
However, even after taking several such uncertainties into account, we find that $\dot{\theta}_{0}$ could be 
constrained at the $10$-$50$~km level, which is still much better than current Solar System constraints and even LISA. 
Moreover, even when accounting for such uncertainties, coincident GW/GRB measurements could constrain a much larger 
region of the parity violating phase space all along the light-cone in which GWs propagate. 

The remainder of this paper presents additional details that lead to the results described above.  
We begin with a description of CS gravity and amplitude birefringence in Sec.~\ref{CS-basics} as a warm-up for the introduction 
of generic gravitational parity violation and its GW observable in Sec.~\ref{gen-amp-bire}. We then proceed with a somewhat
idealized test in Sec.~\ref{Simple-test}, where we compute the modified SNR for a parity-violating signal. Assuming a perfect
electromagnetic and GW determination of the luminosity distance, we then compute the constraint one could place
on parity violation, assuming the latter can be treated as a linear correction to GR. In CS gravity, however, certain resonances
arise in the GW signal at the non-linear level when parity violation is allowed to be strong. Such resonances greatly enhance 
the parity-violating corrections to the SNR, and thus, allow one to place the much stronger constraints shown in Fig.~\ref{fig:intro}, 
as we show explicitly in Sec.~\ref{res-mag}. One might worry that resonances could signal the presence of instabilities, 
but we show that this is not the case in the Appendix. The proposed test, however, is rather idealized because 
correlations with other parameters and astrophysical uncertainties generically degrade measurements of the luminosity distance. 
These potential sources of error are discussed in Sec.~\ref{Ext-disc} and combined with the idealized analysis to construct realistic 
estimates of the possible constraints one could place on parity violation. 

In the rest of this paper we employ geometric units, where the speed of light 
and Newton's gravitational constant are set to unity $c = 1 = G$. 
We further follow the conventions of Misner, Thorne and Wheeler~\cite{Misner:1974qy}, 
where the metric signature is $(-,+,+,+)$, spacetime indices are denoted
with Greek letters, spatial indices with Latin letters and abstract tensors are denoted in boldface type.  

\section{Chern-Simons Modified gravity}
\label{CS-basics}

\subsection{CS Basics}

CS gravity postulates the addition of a parity-violating term to the
vacuum Einstein-Hilbert action (see~\cite{Alexander:2009tp} for a recent review) 
such that~\cite{Jackiw:2003pm} 
\begin{equation}
S = \frac{1}{16\pi} \int d^4x\, \sqrt{-g} \left( R + \frac{1}{4} \theta \;
\mathbf{R\,{}^*R} \right),
\label{CS-action}
\end{equation}
where $g$ is the determinant of the spacetime metric, $R$ is the Ricci
scalar, and
\begin{equation}
\mathbf{R\,{}^*R} = \frac{1}{2} R_{\alpha\beta\gamma\delta}
\epsilon^{\alpha\beta\mu\nu} {R^{\gamma\delta}}_{\mu\nu},
\end{equation}
in terms of the Riemann and Levi-Civita tensors $\mathbf{R}$ and
{\boldmath$\epsilon$}. Such a term is explicitly parity violating because after a spatial
triad inversion, the Levi-Civita tensor changes sign, while the square of the Riemann does not. 
Therefore, the CS correction to the action, the second term in Eq.~\eqref{CS-action}, also changes
sign under parity inversion, provided $\theta$ is a scalar field, i.e.,~it is even under parity.

The CS field equations add terms involving {\boldmath$\nabla$}$\theta$ and 
{\boldmath$\nabla^{2}$}$\theta$ to the Einstein field equations:
\begin{equation}
{\mathbf{G}} + {\mathbf{C}} = 8 \pi {\mathbf{T}},
\label{mod-field-eq}
\end{equation}
where {\boldmath{$G$}} is the Einstein tensor, {\boldmath{$T$}} is the stress energy 
tensor and {\boldmath{$C$}} is the C-tensor
\begin{equation}
C^{\alpha \beta} = \left(\nabla_{\gamma}\theta\right) \epsilon^{\gamma \delta \rho (\alpha}\nabla_{\rho} R^{\beta)}{}_{\delta}
+ \left(\nabla_{\gamma} \nabla_{\delta}\theta\right) {\,^\ast\!}R^{\delta (\alpha \beta) \gamma}\,,
\end{equation}
and parenthesis stand for index symmetrization.
The modified field equations are this simple because the CS correction can be thought of
as a boundary term that acquires dynamics due to the scalar field coupling~\cite{Jackiw:2003pm}. 

CS gravity can lead to two somewhat distinct theories, depending on whether the CS scalar field is
allowed to evolve dynamically or not. 
In what we shall call ``canonical'' or ``non-dynamical''  CS gravity~\cite{Jackiw:2003pm} $\theta$ is
prescribed externally, and a common choice (``static CS gravity'') is to let
{\boldmath$\nabla$}$\theta$ be a constant vector pointing in the direction of
time defined by cosmic expansion.
In what we shall call ``dynamical'' CS gravity~\cite{Gukov:2003cy,
Alexander:2004xd, Smith:2007jm} $\theta$ is promoted to a dynamical field by
adding a kinetic term and a potential $V(\theta)$ to the action, leading to the
equation of motion
\begin{equation}
\square \theta + \frac{dV}{d\theta} = - \frac{1}{4} \mathbf{R{}^*R}.
\label{EOM}
\end{equation}
For the action in Eq.~\eqref{CS-action} to remain invariant under the shift symmetry $\theta \to \theta + c$, where $c$ is a
constant, the potential must vanish, rendering $\theta$ a massless field. In addition to this, 
the dynamical theory also adds a stress-energy tensor for $\theta$ to the right-hand side of Eq.~\eqref{mod-field-eq}.

We consider here both canonical and dynamical CS gravity. In the case of the former, we let $\theta$
be an arbitrary function of time, such that it possesses the same symmetries as a 
Friedman-Roberston-Walker (FRW) spacetime (homogeneity and isotropy). 
In the case of the latter, the $\theta$ field must satisfy its equation of motion, i.e.,~Eq.~\eqref{EOM}, 
whose solution contains a homogeneous and an inhomogeneous piece. The inhomogeneous solution 
is sourced by $\mathbf{R{}^*R}$, which is an extremely small quantity of 
${\cal{O}}(h^{2})$~\cite{Sopuerta:2009iy} in the wave- or radiation-zone 
(many gravitational wavelengths away from the source), where $h$ is the norm of the metric perturbation. 
Therefore, the inhomogeneous solution is also small at the location of the observer and, in fact, parity odd
because it must possess the same symmetries as $\mathbf{R{}^*R}$. Since such a parity odd $\theta$ 
would lead to a parity preserving correction, as then the combination $\theta \mathbf{R{}^*R}$ is parity 
even, we neglect the inhomogeneous solution henceforth. On the other hand, the homogeneous solution
forces $\theta$ to be a freely propagating wave, but since we require $\theta$ to be
homogenous and isotropic, so that the background symmetries are satisfied, we shall assume it is a 
free-function of time only.

At this junction, let us make a brief detour to discuss the the units used for the scalar field.
As written above, $\theta$ has units of squared length, and its time derivative
$\dot{\theta}$ defines a length scale (or an inverse energy or mass scale in natural units), 
which is proportional to the coupling strength of the CS term. To make contact with standard 
notation~\cite{Alexander:2009tp}, $\theta$ is here equivalent to $(16 \pi G) \theta \alpha^{2}/\beta$ 
in the dynamical case or $(16 \pi G) \theta \alpha$ in the canonical or non-dynamical case, where 
$(\alpha,\beta)$ are CS coupling constants (see eg.~\cite{Alexander:2009tp}).

\subsection{Gravitational Waves in CS gravity} 

Consider a circularly-polarized, plane GW, propagating on a flat Minkowski spacetime, of the form
\begin{equation}
h_{\RL}(t,x^i) = h_{0,_{\RL}} \, \exp \left\{ -i \left[ \phi(t) - k_{i} x^i \right] \right\},
\label{wave-ansatz}
\end{equation}
where $t$ and $x^{i}$ are time and space coordinates, $k^{i}$ 
is the spatial wave vector and $h_{0,\RL}$ is some time-independent amplitude. Alternatively, one can 
work directly with the Fourier transform of the waveform
\begin{equation}
h_{\RL}(t,x^{i}) = \frac{1}{(2 \pi)^{3/2}} \int dk \; h_{\RL,k}(t) e^{-i k_{i} x^{i}}\,,
\end{equation}
where we assume the Fourier modes take the form $h_{\RL,k}(t) = h_{0,\RL} e^{-i \phi(t)}$.

After linearizing the CS modified field equations and using 
Eq.~(\ref{wave-ansatz}), the evolution equation for the phase in a right/left circularly-polarized 
basis becomes 
\begin{equation}
\label{wave2}
i\ddot{\phi} + \dot\phi^2 - k^2 = i \lamRL  \frac{k \; \ddot\theta}{1 - \lamRL k \; \dot\theta} \, \dot\phi \,,
\end{equation}
where overhead dots stand for time derivatives, $k \equiv (k_{i} k^{i})^{1/2}$ is the wave number and
where $\lambda_{\R} = +1$ and $\lambda_{\LL} = -1$. Clearly, the phase associated with the right and left polarization
obeys different evolution equations. Formally, the phase should also have a subscript $\RL$, but we drop it
for now to avoid notational clutter. 

Consider now a circularly-polarized, plane GW, propagating on a Friedmann-Robertson-Walker spacetime, of the form
\begin{equation}
\label{wave-ansatz2}
h_{\RL}(\eta,\chi^{i}) = h_{0,\RL} \, \exp \left\{ -i \left[ \phi(\eta) - \kappa_{i}
\chi^{i} \right] \right\},
\end{equation}
where $\eta$ and $\chi^{i}$ are the conformal time and space coordinates,  
and $\kappa^{i}$ is the conformal wave vector. We here let $\phi(\eta)$ be a complex 
function, such that its imaginary part leads to an amplitude time-dependence. Just as before, 
we can alternatively work directly with the Fourier transform of the circularly-polarized waveforms
\begin{equation}
\label{conf-time-FD}
h_{\RL}(\eta,\chi^{i}) = \frac{1}{(2 \pi)^{3/2}} \int d\kappa \; h_{\RL,\kappa}(\eta) e^{-i \kappa_{i} \chi^{i}}\,,
\end{equation}
where now the Fourier modes take the form $h_{\RL,\kappa}(\eta) = h_{0,\RL} e^{-i \phi(\eta)}$.

Once more, after linearizing the modified field equations
in a left/right circularly-polarized basis one finds that the phase obeys a similar 
equation~\cite{Alexander:2004wk, Alexander:2007kv}
\begin{equation}
\label{wave}
i\phi'' + \left(\phi'\right)^2 - \kappa^2 =
-2 i \, \frac{{\cal{H}} - \lambda_{\RL} \kappa \theta''/(2 a^{2})}{1 - \lambda_{\RL} \kappa \theta'/a^{2}} \, \phi'
\end{equation}
where ${\cal{H}} = a'/a$ is the conformal Hubble parameter, with $a$
the dimensionless scale factor that relates conformal and physical time ($dt=a \; d\eta$),
primes are conformal time derivatives and $\kappa \equiv (\kappa_{i} \kappa^{i})^{1/2}$ is the conformal wave number.
One can easily show that Eq.~\eqref{wave} reduces to Eq.~\eqref{wave2} in the limit $a(\eta) \to {\rm{const.}}$

In the flat spacetime case, the modified phase can be obtained by perturbatively solving 
Eq.~(\ref{wave2}): expand the phase as 
\begin{equation}
\label{pert-phase-exp}
\phi = \bar\phi + \delta\phi\,,
\end{equation} 
with $\delta \phi \ll \bar{\phi}$ and assume that $\ddot{\phi} \ll \dot\phi^{2}$ and 
$\delta \ddot{\phi} \ll \dot{\bar\phi} \delta \dot \phi$. The GR solution then returns 
$\dot{\bar{\phi}} = \pm k$. For weak CS parity violation ($k \dot{\theta} \ll 1$ and $\ddot{\theta} \ll 1$), 
the correction to the phase becomes
\begin{equation}
\delta \phi_{\FLAT} = i \lamRL \pi f \int_{t_{s}}^{t_{0}}  \; \ddot{\theta} \, dt\, 
= i \lamRL \pi f  \; \left( \dot{\theta}_{0} - \dot{\theta}_{s}\right)\,,
\label{deltaphiflat2}
\end{equation}
where we have defined the GW frequency $f \equiv k/(2 \pi)$ and
where the subscript ``0'' and ``s'' mean the value at the observer
and the value at the source respectively. Equation~\eqref{deltaphiflat2} reduces to
\begin{equation}
\delta \phi_{\FLAT} = i \lamRL \pi f_{0} D \; \ddot{\theta}_{0},
\label{deltaphiflat}
\end{equation}
if one freezes $\ddot\theta = \ddot\theta_{0} = {\textrm{const.}}$, where $D$ is the comoving distance to the source. 
Notice that it is the combination $D\,f$ that determines the magnitude of the modification, as already discussed 
in the Introduction. 

In the cosmological case, the CS modification to the phase can be obtained in the same fashion
as in the flat case. Expand the phase as in Eq.~\eqref{pert-phase-exp}, but now $\bar{\phi}$ and $\delta \phi$
are functions of conformal time and allowed to be complex. Use this phase ansatz to series-expand 
Eq.~\eqref{wave}, assuming that $\delta \phi \ll \bar{\phi}$, $\phi'' \ll (\phi')^{2}$ and
$\delta \phi'' \ll \bar{\phi}' \delta \phi'$, which follows from $\kappa\theta'\ll1$.
The GR solution then returns ${\bar{\phi}}' = \pm \kappa - i {\cal{H}}$. 
For weak CS parity violation, the correction to the phase becomes
\begin{eqnarray}
\label{deltaphi}
\delta\phi_{\FRW} &=&  i  \lamRL \pi  \int_{t_{s}}^{t_{0}} f(t) \left(\ddot{\theta} - H \dot{\theta}\right) dt\,,
 \\ \nonumber 
&=&  i  \lamRL \pi  a_{0} f_{0} \left( \frac{\dot{\theta}_{0}}{a_{0}} - \frac{\dot{\theta}_{s}}{a_{s}}\right)\,,
\end{eqnarray}
where we have transformed back to cosmic time using $\theta'' = a^{2} (\ddot{\theta} + H \dot{\theta})$, with the
Hubble parameter $H \equiv \dot{a}/a = a {\cal{H}}$. Notice that the GW frequency is formally time-dependent due to the
Hubble dilution, $2 \pi f(t) = k - i H(t)$, but the second term is subdominant and can be neglected;
cf. Section \ref{gen-amp-bire} and particularly the Appendix
[Eq.~\eqref{mainwaveeq}]. 
Also notice that if $\dot{\theta} = 0$ or $H = 0$, we recover the flat spacetime result in Eq.~\eqref{deltaphiflat}. 
Simplifying Eq.~\eqref{deltaphi} by Taylor expanding $\dot{\theta} = \dot{\theta_{0}} + \ddot{\theta}_{0} t$
and using that $a_{s} \simeq a_{0} (1 - z)$ we find
\begin{equation}
\delta\phi_{\FRW} \sim  - i  \lamRL \pi f_{0} z \left(\dot{\theta}_{0} - \frac{ \ddot{\theta}_{0} }{H_{0}} \right)\,,
\label{Tay-Exp-theta}
\end{equation}
where  $H_{0} = (\dot{a}/a)_{0}$ is the value of the Hubble parameter today and
we have assumed a matter-dominated Universe in an $z\ll1$ expansion, suitable for GW/GRB event. 
The derivation presented is consistent with the results found in~\cite{Alexander:2007kv}, 
if one expands the latter in the low-redshift limit. This expansion is justified since 
most LIGO-Virgo GW/GRB events are expected to have $z \ll 1$~\cite{Abadie:2010cf}.

\subsection{Amplitude Birefringence \\ and Parity Violation in CS gravity} 

The imaginary nature of the CS modification to the phase leads to amplitude birefringence, 
i.e.,~the amplitudes of $h_{\LL}$ and $h_{\R}$ are modified differently leading to a suppression/enhancement
with propagation. For weak parity-violation, $\delta \phi \ll \bar{\phi}$, the exponential suppression/enhancement 
of the right/left circularly-polarized GW becomes 
\begin{equation}
h_{\RL} = \hat{h}_{\RL} e^{-i \, \delta \phi_{\RL}} \sim \hat{h}_{\RL} \left(1 - i \, \delta \phi_{\RL}\right),
\label{PV-hRL}
\end{equation} 
where $\hat{h}_{\RL}$ is the GR prediction for the right/left circularly-polarized GW and $\delta \phi_{\RL}$ is
given in Eq.~\eqref{deltaphiflat} or~\eqref{deltaphi}. Notice that we have now reinstated the subscript $\RL$ in the
phase.
Clearly, the parity-violating modification introduces a change in the amplitude
since $i \, \delta \phi_{\RL}$ is a real number.
One should note here that the correction leads to {\emph{finite}} CS corrections, as $\delta \phi_{\RL}$
has all along been assumed small. In the Appendix we show that Eq.~\eqref{PV-hRL} does not signal the presence of an instability
and that indeed the CS correction is formally bounded from above for all relevant values of the CS parameters. 

The degree of birefringence can be quantified through $v$, introduced in Eq.~(\ref{matrix}). 
This equation can be rewritten as $h_{\RL} = \hat{h}_{\RL} (1 + \lambda_{\RL} v)$, which can be inverted
for $v$ to find
\begin{equation}
v = \frac{1}{2} \left(\frac{h_{\R}}{\hat{h}_{\R}} - \frac{h_{\LL}}{\hat{h}_{\LL}} \right) = \frac{i}{2} \left(\delta\phi_{\LL} - \delta \phi_{\R} \right)\,. 
\label{v-def}
\end{equation}
For weak CS parity violation, we have 
\begin{equation}
\label{eq:v:Simple2}
v_{\FLAT,\FRW}  = \pi f  \; \delta\dot{\theta}_{\FLAT,\FRW}\,,
\end{equation}
where we have defined $\delta\dot{\theta}_{\FLAT} = \dot{\theta}_{0} - \dot{\theta}_{s}$ for GWs propagating on flat space
and  $\delta \dot{\theta}_{\FRW} = \dot{\theta}_{0}/a_{0} - \dot{\theta}_{s}/a_{s}$ for GWs propagating on a cosmological background.
To measure this effect, it is crucial to know what amplitude the GW signal
would have in GR, which is very difficult with ground-based detectors alone, 
except in the case of binary inspirals associated with GRB.

The analysis presented so far has ignored the pole in the modified dispersion relation of Eqs.~\eqref{wave2} and~\eqref{wave}
by assuming {\emph{weak}} CS parity violation. This assumption allows us to expand the pole under $k \dot{\theta} \ll 1$. If this assumption
does not hold, then {\emph{strong}} CS parity violation leads to a resonance in the GW response function that might occur inside the sensitivity 
band of the GW detector. This resonance enhances the constraints one could place on gravitational parity violation, as we show 
separately in Sec.~\ref{res-mag} and in the Appendix.

\section{Generic Parity Violation in GW propagation}
\label{gen-amp-bire}

Consider a circularly-polarized, plane GW propagating on an FRW background. 
Let us Fourier decompose this wave as in Eq.~\eqref{conf-time-FD}.
In GR, the linearized field equations give the wave equation
\begin{equation}
\label{waveh}
h''_{\RL,\kappa} + \kappa^2 \, h_{\RL,\kappa} =- 2 \, \mathcal{H} \, h'_{\RL,\kappa}\,.
\end{equation}
The first time derivative of the metric perturbation serves as a dissipative term that 
controls how the wave amplitude drops adiabatically as the universe expands 
when $\mathcal{H}$ is non-vanishing. 

Parity violation leads to a polarization-dependent modification to the amplitude evolution, and thus, 
the parity violating generalization of Eq.~\eqref{waveh} is the replacement of 
\begin{equation}
\label{replacement}
\mathcal{H} \equiv \frac{a'}{a} \to \frac{\mathcal{S}_{\RL}'}{\mathcal{S}_{\RL}}\,,
\end{equation} 
where $\mathcal{S}_{\RL}$ is a polarization-dependent, generalized scale factor. For example, 
In CS theory, the linearized field equations reduce to [see e.g.,~Eq.~$(18)$ in~\cite{Alexander:2004wk}]
\begin{equation}
h''_{\RL,\kappa} + \kappa^{2} \, h_{\RL,\kappa} =-  2 \left[\frac{{\cal{H}} 
- \lambda_{\RL} \kappa \theta''/(2 a^{2})}{1 - \lambda_{\RL} \kappa \theta'/a^{2}} \right] \, h'_{\RL,\kappa}\,
\label{waveh3}
\end{equation}
where we see this is identical to Eq.~\eqref{waveh} with the replacement of Eq.~\eqref{replacement} 
and $\mathcal{S}_{\RL} = a\sqrt{1 - \lambda_{\RL} \kappa \theta'/a^{2}}$.

Solutions to the wave equation can be obtained by assuming a GW ansatz of the form
$h_{\RL,\kappa}(\eta) = e^{-i \phi(\eta)}$, where $\phi$ is  allowed to be complex. In GR, 
this reduces Eq.~\eqref{waveh} to
\begin{equation}
(\phi')^2  - \kappa^2 = - 2  \, i \, \mathcal{H} \, \phi'\,,
\end{equation}
where again we assume $\phi'' \ll (\phi')^{2}$.
For GW wavelengths short compared to the Hubble scale, i.e.,~$\kappa \gg
\mathcal{H}$, the solutions are $\phi' = \pm \kappa - i\mathcal{H}$ to leading
order in $\mathcal{H}/\kappa$ separately for the real and imaginary parts. 
We are led then to the plane-wave GR solution $\hat{h}_{\kappa}$ 
\begin{equation}
\hat{h}_{\RL,\kappa}(\eta) = \frac{a(\eta_{s})}{a(\eta)} \; e^{\pm i \kappa\eta}\,, 
\end{equation}
which are the same irrespective of the GW polarity.

The amplitude-birefringent solutions are obtained similarly through Eq.~\eqref{replacement}. 
The complex phase then becomes 
\begin{equation}
\phi_{\RL} = \pm \kappa \eta - i \int_{\eta_{s}}^{\eta} \frac{\mathcal{S}_{\RL}'}{\mathcal{S}_{\RL}} d\eta'
= \pm \kappa  + i  \ln \left[\frac{\mathcal{S}_{\RL}(\eta_{s})}{\mathcal{S}_{\RL}(\eta)}\right]\,.
\end{equation}
This results in a plane-wave solution of the form \cite{Alexander:2004wk}
\begin{equation}
h_{\RL,\kappa}(\eta) = \frac{\mathcal{S}_{\RL}(\eta_{s})}{\mathcal{S}_{\RL}(\eta)} e^{\pm i \kappa \eta}\,.
\label{exactsol}
\end{equation} 
This solution is subject to the assumption $\kappa \gg \mathcal{S}_{\RL}'/\mathcal{S}_{\RL}$, 
which is satisfied in CS gravity provided $\theta'$ varies on scales
much longer than a gravitational wavelength, i.e.,~$\theta''/\theta' \ll
\kappa$. Of course, this assumption does not hold in the immediate neighborhood 
of the singular point of the wave equation at $\mathcal{S}_{\RL} = 0$.
This singular point is the source of the pole found in the previous section and it is
addressed separately in Sec.~\ref{res-mag} and in the Appendix.

We can now explicitly write the propagator in the form of Eq.~(\ref{matrix}). 
GW polarizations at conformal time $\eta$ are given in terms of $\eta_s$ near
the source as in Eq.~\eqref{matrix} with the parity violating measure
\begin{equation}
v = \frac{1}{2} \frac{a(\eta)}{a(\eta_{s})} \left[ \frac{\mathcal{S}_R(\eta_s)}{\mathcal{S}_R(\eta)} - 
\frac{\mathcal{S}_L(\eta_s)}{\mathcal{S}_L(\eta)} \right].
\label{generalv}
\end{equation}
We note that $v$ reduces exactly to the definition introduced in Eq.~\eqref{v-def} when one sets
$\mathcal{S}_{\RL} = a\sqrt{1 - \lambda_{\RL} \kappa \theta'/a^{2}}$.

\section{A GW/GRB test of weak parity violation}
\label{Simple-test}

Let us now derive the effect of amplitude birefringence on the GW SNR of a GW/GRB event, 
as summarized by~\cite{Dalal:2006qt} and references therein.
The single-detector GW power SNR $\rho$ after matched filtering is given by the inner-product
\begin{equation}
\label{srd}
\rho^2 = \left( h | h \right) \equiv 4 \int_0^\infty df \frac{|\tilde{h}(f)|^2}{S_h(f)},
\end{equation}
where $S_h$ is the one-sided power spectral strain noise density of the detector.
The quantity $\tilde{h}(f)$ is the Fourier transform of the GW strain measured by a detector,
i.e.,~the response function $h(t)$, which in a parity-preserving theory like GR is given by 
\begin{equation}
\label{h}
\hat{h}(t) = F_+ \hat{h}_+(t) + F_\times \hat{h}_\times(t),
\end{equation}
where $\hat{h}_{+,\times}$ are the GR plus or cross-polarized waveform, while
$F_+$ and $F_\times$ are beam pattern functions that depend on the sky
position and a polarization angle $\psi$ (which does not concern us here since
we deal with nearly circular polarization). We use overhead hats
to remind ourselves that these expressions are valid in GR only. In the Fourier domain, 
the GW power of a GR signal is approximately
\begin{equation}
|\hat{\tilde{h}}(f)|^2 = F_+^2 |\hat{\tilde{h}}_+|^2 + F_\times^2 |\hat{\tilde{h}}_\times|^2,
\end{equation}
where $\hat{\tilde{h}}_{+,\times}$ are the Fourier transform of the GR plus or cross-polarized waveform. 
We have neglected the $F_{+} F_{\times}$ term here, since it vanishes upon integration
in the SNR calculation.

For GW signals from binary NS inspirals observed by ground-based
detectors, GR predicts that the frequency-domain signal is given to a good approximation by
\begin{equation}
\hat{\tilde{h}}(f) = \hat{\mathcal{A}} \, f^{-7/6} e^{i\Psi(f)},
\end{equation}
where the phase $\Psi(f)$ does not concern us here (except indirectly as it
yields precise mass measurements which help determine the distance).
We are interested in the amplitude coefficient
\begin{equation}
\hat{\mathcal{A}} = \sqrt{\frac{5}{96\pi^{4/3}}} \frac{\mathcal{M}^{5/6}}{D} \hat{Q},
\label{A-const}
\end{equation}
where $\mathcal{M}$ is the ``chirp mass'' and $D$ is the distance from
source to detector. We neglect corrections to the 
waveform due to propagation on a cosmological background, as most GW/GRB events
will be at low redshift. 
One can encapsulate the sky position, polarization, and
inclination dependence in the factor
\begin{equation}
\label{eq:def:Q}
\hat{Q}^2 = F_+^2 \left( 1 + \xi^2 \right)^2 + 4F_\times^2 \xi^2,
\end{equation}
where $\xi$ is the cosine of the inclination angle between the orbital
angular momentum and the line of sight from source to detector.

A GR binary without orbital precession has a squared SNR of
\begin{equation}
\label{rho2}
\hat{\rho}^2 = \frac{5}{24\pi^{4/3}} \frac{\mathcal{M}^{5/3}}{D^{2}} \hat{Q}^2  I_7\,,
\end{equation}
and thus an SNR of 
\begin{eqnarray}
\hat{\rho} &\approx& 8.3 \left(\frac{{\cal{M}}}{1.2 \; M_{\odot}}\right)^{5/6} \left( \frac{300 \; {\rm{Mpc}}}{D} \right) 
\nonumber \\
&\times&
\left(\frac{I_{7}}{4.7 \times 10^{44} \; {\rm{s}}} \right)^{1/2} \left( \frac{F_{+}^{2} + F_{\times}^{2}}{2/5} \right)^{1/2}\,,
\end{eqnarray}
where in this last relation we have averaged over angles and assumed an inclination angle such that $\xi^{2} = 1$. 
The noise integrals are defined as
\begin{equation}
I_p = \int_0^\infty df \frac{f^{-p/3}} {S_h(f)}.
\end{equation}
Note that $\hat{Q}$, and therefore $\hat{\rho}$, is maximized for $\xi^2=1$, 
i.e.,~looking down the angular momentum axis.

In the limit of strong signals and weak Gaussian noise, the standard
deviation of a single-detector GW amplitude measurement is
\begin{equation}
\frac{\sigma_\mathcal{A}}{\mathcal{A}} = \frac{1}{\rho}.
\end{equation}
With a GW measurement alone it is difficult to convert this into a distance measurement
because sky position, $\xi$ and $\psi$ are not easily measured for
brief signals such as inspirals.
A GW/GRB coincidence provides the needed information.
GRBs with observed afterglows have negligible sky position errors.
If the GRB jet is tightly collimated, $\xi^2=1$ and $\psi$ is irrelevant.
In that case, the best-case GW distance measurement error
\begin{equation}
\label{sigmaD}
\frac{\sigma_D}{D} = \frac{1}{\rho}
\end{equation}
is achievable~\cite{Dalal:2006qt}. Recently,~\cite{Nissanke:2009kt} pointed out that
the above might be an overestimate of the accuracy due to degeneracies between $D$ and $\xi^{2}$.
However these degeneracies can be lifted by observing numerous GW/GRB events
with a network of non-aligned detectors such as LIGO-Virgo.

The single-detector GW SNR is altered in parity-violating gravity as follows:
Combining Eqs.~(\ref{matrix}) and~(\ref{h}), the strain observed by a
detector is seen to be
\begin{equation}
\tilde{h} = F_+ (\hat{\tilde{h}}_+ + i \, v \, \hat{\tilde{h}}_\times) + F_\times (\hat{\tilde{h}}_\times -  i\,  v \, \hat{\tilde{h}}_+)\,.
\end{equation}
and the GW power in the Fourier domain is 
\begin{equation}
|\tilde{h}|^2 = |\hat{\tilde{h}}|^2 + 2 \, v \left( F_+^2 + F_\times^2 \right) 
\mathrm{Im}(\hat{\tilde{h}}_\times\hat{\tilde{h}}_+^*) + {\cal{O}}(v^{2}),
\end{equation}
where again we have neglected the cross terms of the beam pattern functions, as these vanish upon integration. 
This relation then changes $\rho^2$ from the GR value $\hat{\rho}^{2}$ by the replacement
\begin{equation}
\hat{Q}^2 \to \hat{Q}^2 + 4 \, v \left( F_+^2 + F_\times^2 \right) \xi \left( 1 +
\xi^2 \right)  + {\cal{O}}(v^{2}),
\end{equation}
in Eq.~\eqref{rho2}, with $\hat{Q}^{2}$ given by Eq.~\eqref{eq:def:Q}.

Assuming that $v\ll1$, we find that the  change in relative SNR due to weak parity violation can be expressed in terms
of geometrical factors times a weighted average of the model-dependent factor $v$:
\begin{eqnarray}
\label{eq:amp:linear}
\frac{\rho^2}{\hat{\rho}^2} &=& 1 + 4 \,  \left<v\right>  \frac{\left( F_+^2 + F_\times^2 \right)
\xi \left( 1 + \xi^2 \right)} {F_+^2 \left( 1 + \xi^2 \right)^2 +
4F_\times^2 \xi^2}, \\
\label{v-def-ave}
\left<v\right>&\equiv& \frac{1}{I_7} \int_0^\infty df \frac{vf^{-7/3}}{S_h(f)}.
\end{eqnarray}
For a perfectly collimated GRB ($\xi^2=1$) this reduces to the sky-position-independent first-order result
\begin{equation}
\frac{\rho}{\hat{\rho}} \simeq 1 +  \lambda_{\RL} \left<v\right>,
\label{new-rho-eq}
\end{equation}
where the factor of $\lambda_{\RL}$ arises because $\xi=\pm1$ depending on the orientation of the face-on source 
and we have associated a source with $\xi = +1$ with a right-polarized wave.
For a weak CS modification to gravity in an FRW background, 
$v$ reduces to Eq.~\eqref{eq:v:Simple2} and the SNR correction reduces to
\begin{equation}
\label{eq:drho:WeakCS2}
\frac{\rho}{\hat{\rho}} = 1 + \lambda_{\RL} \pi \; \delta \dot{\theta}_{\FRW} \; \frac{I_4}{I_7}\,,
\end{equation}
As in Eq.~\eqref{Tay-Exp-theta}, in the limit of slowly-varying $\dot{\theta}$, Eq.~\eqref{eq:drho:WeakCS2} reduces to
\begin{equation}
\label{eq:drho:WeakCS}
\frac{\rho}{\hat{\rho}} = 1 - \lambda_{\RL} \pi \, z \, \left(\dot\theta_0 - \frac{\ddot{\theta}_{0}}{H_{0}} \right) \frac{I_4}{I_7}\,,
\end{equation}
 where we have used that $D \approx z/H_{0}$ for $z \ll 1$. 

In the best case, $\rho$ is a Gaussian with variance unity, and thus 
a $(1\sigma)$-bound on generic parity violation translates into 
\begin{equation}
\left<v\right> < 1/\rho\,.
\end{equation}
For weak CS parity violation with $\dot{\theta} = \dot{\theta}_{0}$, this test would allow one to place the constraint
\begin{equation}
\label{eq:constraintEqn:WeakCS}
|\dot\theta_0| < \frac{1}{\pi} \frac{I_7}{I_4} \frac{1}{z \, \rho}\,.
\end{equation}
With Eq.~(\ref{rho2}), Eq.~(\ref{eq:constraintEqn:WeakCS})
can be rewritten in terms of observables as
\begin{equation}
\label{single}
|\dot\theta_0^{-1}| > \sqrt{\frac{5}{6}} \pi^{1/3} \sqrt{ F_+^2 + F_\times^2 } H_0
\mathcal{M}^{5/6} \frac{I_4}{\sqrt{I_7}}.
\end{equation}

For a double NS GW/GRB, averaging over sky location (such that $F_{+}^{2} = F_{\times}^{2} = 1/5$),
$H_0=70.5$~km~s$^{-1}$~Mpc$^{-1}$, 
and a single interferometer with the present estimate of the
``advanced LIGO'' fully-commissioned noise 
performance~\cite{AdvancedLIGONoise,2010CQGra..27h4006H}, the threshold is
\begin{equation}
\label{result}
|\dot\theta_0| < 1200~\mbox{km} \left( \frac{1.2~M_\odot} {\mathcal{M}}
\right)^{5/6} \left( \frac{2/5} {F_+^2 + F_\times^2} \right)^{1/2}.
\end{equation}
Therefore, the 95\% confidence threshold for parity violation (compatible with a $2\sigma$ constraint) 
translates into the bound $|\dot\theta_{0}| < 2400~\mbox{km}$. 
Note that the dependences on $z$ and $\rho$ have cancelled out.
However, $\rho$ does matter in that $z$-independent factors which improve
$\rho$, also improve the threshold on $\dot\theta_{0}$, up to a maximum $\rho$ set by
systematic errors (see Sec.~\ref{Ext-disc}).

\section{Resonant Magnification}
\label{res-mag}

The analysis presented above neglects a resonance  in the parity violating
interaction [see e.g.,~Eqs.~\eqref{wave2},~\eqref{wave} or Eq.~\eqref{generalv}]. 
As is apparent from the ubiquitous denominator $1/{\cal{S}}_{\RL}$ in $v$, 
certain wavelengths will be strongly amplified. Such an amplification leads to a {\emph{finite}} increase in the 
amplitude correction, as it is shown in the Appendix. Without specifying a functional form for ${\cal{S}}_{\RL}$, it is
nearly impossible to study such resonant behavior, which is why in the rest of this section we assume ${\cal{S}}_{\RL}$
is given by the CS prediction in Sec.~\ref{gen-amp-bire}. 
 
In CS gravity, the resonant frequency $f_{p}$ and resonant conformal time $\eta_{p}$ are such that ${\cal{S}}_{\RL} = 0$, 
which reduces to the condition $\kappa_{p} = \lambda_{\RL} a^{2}_{p}/\theta'_{0}$ or in terms of cosmic time
\begin{equation} 
f_{p,0} = \frac{\lambda_{\RL}}{2 \pi \dot{\theta}_{0}}\,. 
\end{equation}
Nearly the same frequency band will be resonant for \emph{all} GW sources, 
because the resonance condition depends only on $\dot\theta_{0}$ at the detector. 

The estimates of Sec.~\ref{gen-amp-bire} for amplification are adequate, provided the GW frequency
is sufficiently far from this present-day  resonance $f_{p,0}$; the resonant frequency at the source $f_{p,s}
=\lambda_{\RL}/2\pi\dot{\theta}_s$; and any frequency in between, as these frequencies have generally been resonant in
the past.   As described in the Appendix, the strongest amplification will occur very close to the present-day resonant
frequency $f_{p,0}$,
with a bandwidth of 
$\delta f_{p} \approx (\dot{f}_{p})^{1/2}$, which reduces to $\delta f_{p} = [\ddot{\theta}/(2 \pi \dot{\theta}^{2})]^{1/2}$. 
Assuming that $\theta$ evolves on cosmological scales $\ddot{\theta} = {\cal{O}}(H \dot{\theta})$ 
implies that $\delta f_{p} = {\cal{O}}[(H/\dot{\theta})^{1/2}] = {\cal{O}}(10^{-8} \; \rm{Hz})$. 

Similarly, the time period during which the signal is resonant can be computed from the inverse of the resonance bandwidth, or
equivalently from the rate of change of the resonant conformal time $\eta_p$ with respect to conformal wavenumber 
$\kappa$: ${\cal{R}} \equiv d \eta_{p}/d\kappa_{p}$. Let us define the squared amplitude 
$\calF_{\RL} \equiv {\cal{S}}_{\RL}^{2}$, which is a function of two independent variables: 
$\eta_{p}$ and $\kappa_{p}$. Requiring that $d\calF_{\RL} = 0$ leads to
\begin{equation}
\frac{d\eta_p}{d\kappa_{p}} = - \frac{d \calF_{\RL}/d\kappa_{p}}{d\calF_{\RL}/d\eta_{p}}\,,
\end{equation}
and rewriting this equation slightly we find 
\begin{equation}
{\cal{R}} = -\frac{d}{d\kappa} \left(\frac{\calF_{\RL}}{\calF'_{\RL}}\right)\,,
\label{R-def}
\end{equation}
where we have used the fact that $\calF_{\RL}=0$ at resonance. 

Although the resonance bandwidth is small, if a resonance is presence in the sensitive band of the GW detector 
it will greatly enhance the constraints of the previous section. In what follows, we compute the enhancement of the simple 
GW/GRB test discussed in Sec.~\ref{Simple-test} if the resonance is in-band. We begin by 
assuming that the GW signal detected is parity violating and one searches over the data with parity-violating
templates. Such considerations lead to an optimal constraint related to the intrinsic SNR of the parity-violating templates. 
We follow this analysis by considering a parity-violating signal that is filtered with a GR template. By doing so, one incurs a fundamental
error in the determination of parameters induced by the assumption that the data contains a GR template (see e.g.,~\cite{Yunes:2009ke} 
for a more detailed discussion of fundamental bias in GW astrophysics). 

\subsection{Optimal Strategy}
\subsubsection{Signal-to-Noise Ratio and an Optimal Test}

The presence of a resonance in the frequency band where searches are performed leads to an amplification of parity violation.
Such an amplification can lead to a stronger constraint than the one we considered in Sec.~\ref{Simple-test}. Of course, the constraint
is largest when the search is performed with templates that account for such parity violation. We consider such cases in this subsection.

Let us then assume that some of the detected frequencies are resonant and that one
correctly identifies all signal power associated with the event to the correct parity-violating template. 
The SNR then becomes\footnote{Technically, the one-sided SNR integral arises from a more general, two-sided overlap 
integral~\cite{CutlerFlanagan:1994}, acting on a real-valued (i.e.,~single polarization) strain signal $h_{+}(t)$. Even though
the $\calF_{\RL}$ is not symmetric in $f$, the Fourier transform can be folded together using  
that $\calF_{\R}(f) = \calF_{\LL}(-f)$.}
\begin{eqnarray}
\rho^2 &\equiv& 4  \int_{0}^\infty df \frac{|\tilde{h}|^2}{S_h(f)} 
= 4  \int_{0}^{\infty} \frac{df}{S_h} \left| \tilde{h}_{\R} F_{\R} + \tilde{h}_{\LL} F_{\LL} \right|^{2}
\\
&=& \int_{0}^{\infty} \frac{df}{S_h} \left\{ 
\left(X_{\R}^{2} + X_{\LL}^{2}\right) \left(|\hat{\tilde{h}}_{+}|^{2} + |\hat{\tilde{h}}_{\times}|^{2} \right) \left(F_{+}^{2} + F_{\times}^{2} \right)
\right. 
\nonumber \\
&+& \left. 2  \left(X_{\R}^{2} - X_{\LL}^{2} \right) \left(F_{+}^{2} + F_{\times}^{2}\right)
 {\rm{Im}} \left(\hat{\tilde{h}}_{\times} \hat{\tilde{h}}_{+}^{*}\right)
\right. 
\nonumber \\
&+& \left. 2 X_{\R} X_{\LL} \left(|\hat{\tilde{h}}_{+}|^{2} - |\hat{\tilde{h}}_{\times}|^{2} \right) \left(F_{+}^{2} - F_{\times}^{2} \right)
\right\}\,,
\label{newResSNR-gen}
\end{eqnarray}
where we have defined $F_{\R,\LL} \equiv (F_{+} + \lambda_{\RL} F_{\times})/\sqrt{2}$ and $X_{\RL} \equiv (\calF_{\RL,s}/\calF_{\RL,0})^{1/2}$, 
and where we have assumed $h_{+}$ and $h_{\times}$ are uncorrelated, so that their integrated product vanishes. 
Recall that hatted quantities correspond to the GR expectation, so that $\hat{\tilde{h}}_{+,\times}$ are the Fourier transform of the
GR, plus or cross-polarized waveform. Notice that this expression reduces
to the GR result when $X_{\R} = X_{\LL} = 1$ and to the expression in Eq.~\eqref{eq:amp:linear} 
in the linear approximation $X_{\RL} \sim 1 + \lambda_{\RL} v$. 

The above expression can be simplified somewhat if we assume the source is at $\xi = \pm 1$. 
We then find 
\begin{eqnarray}
\rho^2 &=& \frac{5}{12 \pi^{4/3}} \frac{{\cal{M}}^{5/3}}{D^{2}} \left(F_{+}^{2} + F_{\times}^{2}\right) 
\nonumber \\
&\times&
\int_{0}^{\infty} \frac{df}{S_h} f^{-7/3} \left[ \left(X_{\R}^{2} + X_{\LL}^{2}\right)
\pm \left(X_{\R}^{2} - X_{\LL}^{2} \right) \right]\,.\;
\label{newResSNR-gen2new}
\end{eqnarray}
The beam-pattern functions then only modify the overall coefficient of the SNR, and thus 
we can rewrite the SNR in the more familiar form 
\begin{equation}
\rho^2 = 2 {\cal{A}}^{2} \int_{0}^{\infty} \frac{df}{S_h} f^{-7/3} \left[
\left(X_{\R}^{2} + X_{\LL}^{2}\right) 
\pm   \left(X_{\R}^{2} - X_{\LL}^{2} \right) \right]\,,
\label{newResSNR-gen2}
\end{equation}
Given a single detector, we cannot distinguish between a purely right, or purely left-polarized signal, so 
effectively we measure either $\rho_{\R}$ or $\rho_{\LL}$, where we have defined
\begin{equation}
\label{newResSNR}
\rho^2_{\RL} = 4 \frac{{\cal{A}}^{2}}{1 + z} \int_{0}^\infty
df \frac{f^{-7/3}}{S_h(f)} \; (1 + z) \;
 \left|  \frac{\calF_{\RL}(\eta_s, k)}{
         \calF_{\RL}(\eta_{0}, k)}
  \right| \equiv \hat{\rho}^2{\calAmp_{\RL}}^2\,,
\end{equation}
and the amplification factor is
\begin{equation}
\label{newResSNR2}
{\calAmp_{\RL}}^2 \equiv \frac{1}{I_7} \int_{0}^\infty df \frac{f^{-7/3}}{S_h(f)} \;
 (1 + z) \; \left|
   \frac{\calF_{\RL}(\eta_s, k)}{
         \calF_{\RL}(\eta_{0}, k)}
  \right|.
\end{equation}
We have here used that $\hat{\rho} = 4 {\cal{A}}^{2} I_{7}/(1 + z)$ as shown in Eq.~\eqref{rho2}, 
but we have included a factor of $(1 + z)^{-1}$ to map between comoving and luminosity distances.
This result generalizes the linear, weak-amplification expression of Eq. (\ref{eq:drho:WeakCS}).  

Let us now show that the above reduces to the results of Sec.~\ref{Simple-test} in the weak parity violation limit. 
Assuming the resonant frequencies are very large (i.e.,~$k_{0} \dot\theta \ll 1$) and that the source
is at low redshift $D H_{0} \ll 1$, the resonant factor becomes to leading order
\begin{equation}
\left| \frac{\calF_{\RL}(\eta_s, k)}{\calF_{\RL}(\eta_{0}, k)}\right| 
   \simeq 1 - 2 \pi f D \lambda_{\RL} \left(\dot{\theta}_0 H_{0} - \ddot{\theta}_0 \right)\,.
\end{equation}
We recognize the right-hand side as simply $1 + 2 \lambda_{\RL} v$ and inserting this back into the SNR we find
\begin{eqnarray}
\rho^{2} &\simeq& \hat{\rho}^{2} \frac{1}{I_{7}} \int_{0}^{\infty} df \frac{f^{-7/3}}{S_{h}(f)} \left(1 + 2 \lambda_{\RL} v \right)\,,
\nonumber \\
&\simeq& \hat{\rho}^{2}  \left(\frac{1}{I_{7}} \int_{0}^{\infty} df \frac{f^{-7/3}}{S_{h}(f)}+ 2 \lambda_{\RL} \frac{1}{I_{7}} \int_{0}^{\infty} df \frac{f^{-7/3}}{S_{h}(f)} v \right)\,,
\nonumber \\
&\simeq& \hat{\rho}^{2}  \left( 1 + 2 \lambda_{\RL} \left<v\right>\right)\,,
\end{eqnarray}
where $\left<v\right>$ was defined in Eq.~\eqref{v-def-ave}.
Notice that this expression reduces to Eq.~(\ref{new-rho-eq}), upon linearizing the square-root.

A test of parity violation then presents itself. Let us assume that a GW detection has been made 
coincident with a GRB event at a known distance, so that the chirp mass and $\hat{\rho}$ (the GR expectation) 
are unambiguously known. Then, parity violation can be distinguished from a
$(2 \sigma)$-statistical fluctuation if $|\rho_{\RL}/\hat{\rho}-1|>2/\hat\rho$ [cf.~Eq.~(\ref{eq:constraintEqn:WeakCS})]. 
If such a violation is not observed, one can rule out points in the $(\dot\theta,\ddot\theta)$ space. In particular, 
$(\dot\theta_{0},\ddot\theta_{0})$ values that lead to 
\begin{equation}
|\calAmp_{\RL} -1|> \frac{2}{\hat{\rho}}\,,
\label{opt-test}
\end{equation}
can be excluded; inversely, $(\dot{\theta}_{0},\ddot{\theta}_{0})$ values that lead to the opposite relation
cannot be distinguished from pure GR with a single observation.

\subsubsection{Numerical Implementation}

The implementation of the test in Eq.~\eqref{opt-test} requires the calculation of the amplitude enhancement 
factor $\calAmp_{\RL}$, which in turn requires knowledge of the ratio of $\calF_{\RL}$ at the source and at the 
observer location. The latter can be rewritten as
\begin{equation}
\label{eq:Amp:Approx}
(1 + z) \; \left| \frac{\calF_{\RL}(\eta_s, k)}{\calF_{\RL}(\eta_{0}, k)}\right| =  \left| \frac{1 - f/(a_{s} f_{p,s})}
{1- f/(a_{0} f_{p,0})} \right|,
\end{equation}
where $a_{0}/a_{s} = 1+z$ and we have defined the resonant frequencies at the source $f_{p,s}$ and at the observer $f_{p,0}$. 
The resonant or pole frequency at the observer is simply $f_{p,0} = \lambda_{\RL}/(2 \pi \dot{\theta}_{p,0})$, while that at the source can be approximated as
\begin{equation}
f_{p,s} = f_{p,0} + \frac{df}{dt} \delta t = f_{p,0} - \frac{D}{2 \pi {\cal{R}}} + f_{p,0} H_{0} D,
\label{fps-Tay-Exp}
\end{equation}
where we have used that $f = \kappa/(2 \pi a)$, $\dot{f}_{p} = \kappa'_{p}/(2 \pi a^{2}) - \kappa_{p} {\cal{H}}/(2 \pi a^{2})$, $\kappa'_{p} = {\cal{R}}^{-1}$ and $\delta t = -D$. 
Combining $f_{p,s}$ with the appropriate redshift factors that arise in $a_{s}\approx (1 - z) a_{0}$, we find
\begin{equation}
a_{s} \, f_{p,s} \approx f_{p,0} - \frac{D}{2 \pi {\cal{R}}} + {\cal{O}}(k_{p,0}^{2}H_{0}^{2}),
\label{fs}
\end{equation}
where the last term of Eq.~\eqref{fs} has cancelled.

The calculation of the enhancement factor then requires knowledge of ${\cal{R}}$, which following Eq.~\eqref{R-def} reduces to
\begin{equation}
{\cal{R}} = \frac{\lambda_{\RL}}{4 H_{0}} \left(\dot{\theta}_{0} - \frac{\ddot{\theta}_{0}}{H_{0}}\right) 
\left[1 - \frac{\lambda_{\RL}k_{p,0}}{2} \left( \dot{\theta}_{0} + \frac{\ddot{\theta}_{0}}{H_{0}} \right) \right]^{-2}\,.
\end{equation}
Using further the fact that $k_{p,0} = f_{p,0}/(2 \pi) = \lambda_{\RL}/(4 \pi^{2} \dot{\theta}_0)$, we can simplify the above expression to
\begin{eqnarray}
{\cal{R}} &=& \frac{1}{2\pi f_{p,0}H_0(1-q)}\,, \\
q &\equiv&  \ddot{\theta}_{0}/(\dot{\theta}_{0} H_{0})\,,
\end{eqnarray}
where the dimensionless parameter $q$ is $O(1)$ for slowly-varying $\dot{\theta}$.
Using this expression and in the limit
$z\approx H_0 D\ll 1$,  Eq.~\eqref{fs} reduces to the extremely simple relation
\begin{eqnarray}
a_{s} \, f_{p,s} \approx f_{p,o}\left[1+z  (1-q)\right]\,.
\end{eqnarray}

In the local universe or for slowly-varying $\dot\theta$, the present-day and source-frame resonant frequencies are nearly identical.
More precisely, the first-order approximation used in Eq.~(\ref{fs}) to relate these frequencies implicitly requires  $z
|(1-q)| \ll 1$, or
\begin{equation}
z |\ddot{\theta}_{0}-H_0 \dot{\theta}_{0}| \ll  |H_0\dot{\theta}_{0}|\,,
\label{gray-region}
\end{equation}
corresponding to the requirement that $\delta \dot{\theta}_{\FRW}$ be small. The analysis presented here does not formally apply 
when the pole frequency is rapidly varying, i.e.,~when Eq.~\eqref{gray-region} is violated. 
However, even then, non-resonant and resonant amplification will generically occur and one should still be able to constrain 
gravitational parity violation. The results we present below, however, are not valid in this region.

The integrals in Eq.~\eqref{newResSNR} and~\eqref{newResSNR2} are evaluated with the estimated NS-NS optimized spectral noise density
of~\cite{AdvancedLIGONoise} and numerically converting them into Riemann sums and the Fourier integrals into
Fourier series. In doing so, we must choose a frequency discretization $df$, which is related to the observation time
$T$ via $df \simeq 1/T$. Full coherent recovery of the amplified waveform clearly corresponds to an unrealistically long 
coherently-integrated data stream: for a $T=1$ year long integration, $df = {\cal{O}}(1/{\rm{yr}}) \simeq 10^{-8}\unit{Hz}$.  
In practice, due to detector dropouts and nonstationarity of the noise, the
longest integration times are 2048~s~\cite{Abbott:2007xi},
imposing a minimum frequency discretization of $d f \simeq 1/T \simeq 5 \times 10^{-4} \; {\rm{Hz}}$.  
A search strategy based on these limited data intervals will always recover 
a somewhat smaller resonant contribution to the total SNR than if one where to integrate for larger $T$, 
which then implies less limiting constraints on $(\dot{\theta}_{0},\ddot{\theta}_{0})$, 
assuming the same SNR threshold.
However the dependence on $df$ is generally quite weak.
We have verified these statements by numerically computing the integrals in Eq.~\eqref{newResSNR} 
and~\eqref{newResSNR2} both with $df = 10^{-8} \; {\rm{Hz}}$ and $df=1/2048$~Hz
and checking that the changes to Fig.~\ref{fig:amp} (described below) are
hardly visible.
In what follows, we shall adopt a realistic frequency resolution of
$df=1/2048$~Hz. 

The integration of the resonant amplitudes are also formally ill-defined when 
the integrand sweeps through the resonance $f \to f_{p,0}$, as then $\calF_{\RL}(\eta_{0},k)$ vanishes. Such divergent behavior
can be cured by regularizing all integrals, i.e.,~by cutting out a region of size $\epsilon$ about the resonant frequency
at present ($f_{p,0}$). Such a regularization is valid because the SNR depends logarithmically on the regulator, as we show explicitly in the Appendix. 
We choose here $\epsilon  = \delta f_{p}$, but we have checked that choosing $\epsilon = df = 10^{-4} \; {\rm{Hz}}$, for example, does not visually
affect any of the plots. This regularization is allowed within the context of linear propagation theory because the resonant amplification 
enhances the waveform only to $h \, f/\delta f_{p}  \ll 1$. Even though the enhancement factor
is formally large, $f/\delta f_{p} \simeq O(10^{10})$, the overall correction is clearly small as typical GW strains 
are $h = {\cal{O}}(10^{-22})$. In the Appendix, we show in more detail that regularizing these integrals induces a negligible error in the SNR. 

\subsubsection{Results and Discussion}

\begin{figure}
\includegraphics[width=\columnwidth,clip=true]{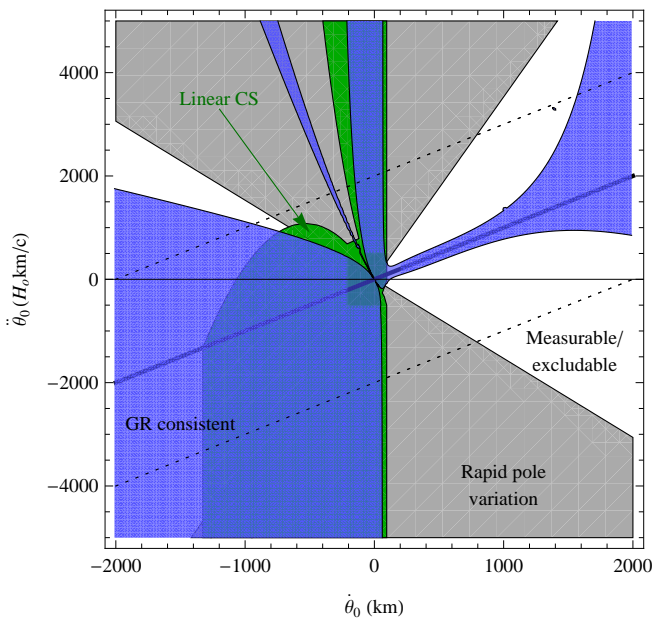} \; \;
\caption{\label{fig:amp}
The blue shaded region corresponds to values of $(\dot{\theta}_{0},\ddot{\theta}_{0})$ for which parity 
violation is too small to be measured with a coincident GRB/GW event. The region outside this blue-shaded area
can be measured or excluded given a single, GW/GRB event at $\hat{\rho} = 10$. The boundaries of the blue-shaded region correspond to the best
$(\dot{\theta}_{0},\ddot{\theta}_{0})$ constraints one could place. 
The thick solid line corresponds to the values of $(\dot{\theta}_{0},\ddot{\theta}_{0})$ for which there is no parity violation ($\calAmp_{\R}=1$), 
and thus it represents an absolute theoretical limit to the constraints one could place on CS amplitude birefringence.
The dotted lines bracket the values of $(\dot{\theta}_{0},\ddot{\theta}_{0})$ that are allowed by Solar System constraints. 
Consequently, the regions outside the dotted lines (not shaded in the figure) have already been excluded by Solar System tests.
The green-shaded region shows where the linear, weak-parity approximation of Sec.~\ref{Simple-test} is valid, 
while the gray-shaded region shows where the linear cosmological expansion of Eq.~\eqref{fs} breaks down. 
Our calculations suggest that LIGO/Virgo will be able to see or rule out the entire 
area outside the blue-shaded region, excluding a much larger region of parameter space than Solar System constraints. 
}
\end{figure}

Figure~\ref{fig:amp} shows numerical results for the implementation of the test devised in Eq.~\eqref{opt-test}.
The blue-shaded region, labeled {\emph{GR consistent}, corresponds to values of $(\dot{\theta}_{0},\ddot{\theta}_{0})$ for which we cannot 
distinguish between GR and CS gravity, i.e.,~values for which $|\calAmp_{\R} -1| < 2/\hat{\rho}=2/10$. 
This region is generated assuming $D=600 \; \unit{Mpc}$ and $\hat{\rho} = 10$ as a fiducial threshold, 
which could occur in advanced LIGO given a NS-NS merger at an optimal sky location. 
Given a coincident GRB/GW detection that is non-parity violating, LIGO/Virgo will be able to observe or rule out the area {\emph{outside}} 
the blue-shaded region in the figure, which is why we have labeled it {\emph{measurable/excludable}}.
Observe that the blue-shaded region in this figure corresponds to the non-shaded region in Fig.~\ref{fig:intro}.

The boundary of the blue-shaded region in Fig.~\ref{fig:amp} denotes the best bound that one could place 
on $(\dot{\theta}_{0},\ddot{\theta}_{0})$ given a coincident GRB/GW detection consistent with general relativity. This boundary corresponds
to values of $(\dot{\theta}_{0},\ddot{\theta}_{0})$ for which $|\calAmp_{\R} -1| = 2/\hat{\rho}=2/10$. 
The Solar System constraint $\dot{\theta}_{0} < 2000 \; {\rm{km}}$ with $\ddot{\theta}_{0} = 0$ is outside the
plotting range of this figure. One can generalize or improve on the Solar System constraint via the map 
$|\dot{\theta}_{0}| \to |\dot{\theta}_{0} - \ddot{\theta}_{0}/H_{0}|$, assuming the entire right-hand side can be treated
as a constant, evolving on timescales much larger than the orbital velocity of the LAGEOS satellites. Doing so,  
the improved Solar system constraint becomes $|\dot{\theta}_{0}-\ddot{\theta}_{0}/H_0|<2000 \; \unit{km}$. 
We have plotted this constraint with two diagonal dotted lines in Fig.~\ref{fig:amp}. The region between 
these dotted lines is consistent with GR and cannot be ruled out, while the area outside of the dotted lines
can be excluded with the improved Solar System constraint. 
  
We can now compare this generalized Solar System constraint (the dotted lines) to the possible GW/GRB constraints 
(the boundary of the blue-shaded region). Figure~\ref{fig:amp} shows that the latter largely is inside the former when $\dot{\theta}_0>0$, 
which implies GW/GRB constraints would be better than the generalized Solar System constraints.
When $\dot{\theta}_0< 0$ the GW/GRB constraint is not as good with a single detector, as CS parity violation would 
then be non-resonant in $\calAmp_{\R}$. Such an issue is circumvented  if either (a)  two GW detectors are used to extract both 
$h_{\R}$ and $h_{\LL}$ for a non-optimally aligned source $|\xi| \ne 1$
(we discuss this in the next paragraph) or (b) two sources, one right- and one left-handed, have been observed. Furthermore, if we assume $\ddot{\theta}_0 = 0$ as in the standard Solar System constraint, 
or more generally whenever $|\ddot{\theta}_0| < 100 \; H_{0} \; {\rm km} \; {\rm{s}}^{-1}$, our results suggest that a GW/GRB detection 
could constrain $\dot{\theta}_{0} < 100 \; {\rm{km}}$. This bound is $20$ times stronger than the current Solar System constraint 
($\dot{\theta}_{0} <  2000 \; {\rm km}$). Such enhancements in the GW constraints relative to those presented in 
Sec.~\ref{Simple-test} are clearly due to the CS resonance in the SNR.    

Resonance amplification for a right-circularly polarized GW only occurs when the integrand of the amplitude factor $\calAmp_{\R}$ contains a pole,
which leads to the logarithmic divergence discussed in the Appendix.  For $\dot{\theta}_{0} < 0$, the amplification of
right-handed signals
is suppressed as the pole moves to negative frequencies;  constraints from right-handed measurements on
$\dot{\theta}_o,\ddot{\theta}_o$ are therefore not as strong. The situation reverses, however, 
if the GW is left-polarized, as then the $\calAmp_\LL$ regions are identical to the $\calAmp_\R$ ones in Fig.~\ref{fig:amp} 
but with $(\dot{\theta}_{0},\ddot\theta_{0}) \rightarrow (- \dot{\theta}_{0},-\ddot{\theta}_{0})$. Therefore, the two-detector detection of a 
a right and a left-polarized GW would allow us to constrain $(\dot{\theta}_{0},\ddot\theta_{0})$ equally well for both positive and negative 
$\dot{\theta}_{0}$. More precisely, with such a detection one would be able to constrain $|\dot{\theta}_0| < 100 \; {\rm{km}}$ for any 
$|\ddot{\theta}_0| < 100 \; H_{0} \; {\rm km} \; {\rm{s}}^{-1}$.

An interesting feature of CS parity violation not appreciated with Solar System constraints is that for certain values of $(\dot{\theta}_{0},\ddot{\theta}_{0})$ 
amplitude birefringence can actually vanish. In some sense, for these values of the derivatives of the CS coupling, the CS correction conspires with the 
cosmological expansion to lead to no parity violation. Such values of $(\dot{\theta}_{0},\ddot{\theta}_{0})$ then constitute the ultimate 
theoretical limit to which CS gravity can be constrained. These values of the derivatives of the CS coupling are such that $\calAmp_{\RL} = 1$, 
which implies $\rho = \hat{\rho}$. This degeneracy occurs when $f_{p,0} = f_{p,s}$ or simply when $\ddot{\theta}_{0} = H_{0} \dot{\theta}_{0}$, 
which is simply a straight line in the $(\dot{\theta}_{0},\ddot{\theta}_{0})$ phase space. We have plotted this linear relationship as a solid diagonal 
line in Fig.~\ref{fig:amp}. Observe that this line is always inside the blue-shaded region, as it should, as the latter represents precisely the region 
where a GW/GRB measurement does not have the sufficient accuracy to see parity violation given an SNR of $10$. 

Although resonant amplification generally occurs for any prescribed form for $\theta$, the Taylor expansion used in Eq.~\eqref{fps-Tay-Exp}
to relate source quantities to observer quantities breaks down for certain values of $(\dot{\theta}_{0},\ddot{\theta}_{0})$. More precisely,
when $(\dot{\theta}_{0},\ddot{\theta}_{0})$ is such that $|z(1-q)| = {\cal{O}}(1)$ or equivalently 
$|H_{0} \dot{\theta}_{0} - \ddot{\theta}_{0}| = {\cal{O}}(H_{0} \dot{\theta}_{0}/z)$, the pole frequency changes rapidly and the 
Taylor expansion of Eq.~\eqref{fps-Tay-Exp} is not valid. One might then worry that our analysis breaks down in this region 
and the bounds quoted earlier in terms of the boundary of the blue-shaded region would not be valid, 
as higher-order terms in Eq.~\eqref{fps-Tay-Exp} would be needed. To get a sense of the $(\dot{\theta}_{0},\ddot{\theta}_{0})$ 
values for which this happens, in Fig.~\ref{fig:amp} we have shaded in gray the region where $|z(1-q)|>0.5$ and labeled it {\emph{rapid pole variation}}. 
Observe that this region does not overlap the blue-shaded region for $\dot{\theta} > 0$, which explicitly shows that the bounds proposed above still hold.
Of course, one can shrink this gray region by Taylor expanding Eq.~\eqref{fps-Tay-Exp} to second-order if needed, at the expense of
introducing a new parameter $\dddot{\theta}_0$.

The numerical results we have presented here do recover the linear approximation [i.e.,~Eqs.~\eqref{eq:drho:WeakCS} and~\eqref{Tay-Exp-theta}]
for the $(\dot{\theta}_{0},\ddot{\theta}_{0})$ values where this approximation is valid, i.e.,~where there is no
resonance in the detection band. The source and observer resonances are out of band provided either $f_{p,0}<-\left<f\right>$ 
and $f_{p,0}>300\unit{Hz}$ (for the suppressed frequency being negative or out of the most 
sensitive band) or $f_{p,s}>500\unit{Hz}$ and $f_{p,s}<-\left<f\right>$ (when the resonance is well out of band or negative). 
In order to visualize this in the context of the bounds placed by the boundary of the blue-shaded region in Fig.~\ref{fig:amp}, we have shaded in green
the linear approximation region and labeled it {\emph{linear CS}}.  Observe that the linear approximation breaks down for
all physically interesting and accessible  $\dot{\theta}_0>0$:  a resonance generically appears inside the LIGO band. Notice that this is not the case for LISA, as this instrument is sensitive to much lower
frequency GWs, for which the resonance is always out of band if one saturates $\dot{\theta}_{0}$ with the Solar system constraint. 

Finally, we have checked that the constraints presented here are formally independent of distance to the source for $(\dot{\theta}_{0},\ddot{\theta}_{0})$
in the linear regime, which is consistent with the observations of Sec.~\ref{Simple-test}. This occurs because although CS amplitude 
birefringence is proportional to distance [see e.g.,~Eq.~\eqref{eq:drho:WeakCS}], the dominant source of error also grows with distance. 
Therefore, to leading order, any single, short GRB observation consistent with pure GR translates into upper limits on $\dot{\theta}_0-\ddot{\theta}_o/H_0$ 
that are \emph{independent} of distance and depend only on the absolute detector sensitivity 
along the line of sight [Eq.~\eqref{result}]\footnote{As in the linear case, the expression $\left<v\right><1/\rho$ misleadingly suggests distance-dependent results. 
Areas in Fig.~\eqref{fig:amp} only apply to  self-consistent choices for $\rho$ and $D$.  For example, the contour for $|\calAmp-1|<2/10$ at 
$D=600\unit{Mpc}$ corresponds to either twice as sensitive an instrument or $2^{6/5}$ times larger a binary chirp mass.}. On the other hand, 
in the {\emph{non-linear}} regime, where resonant behavior is present and the linear approximation is not valid, the blue shaded region in
Fig.~\ref{fig:amp} does depend strongly on source distance.

\subsection{Suboptimal Strategy}

The previous analysis assumed the entire signal, including its narrow resonantly amplified contributions, are correctly
identified through an appropriate, parity-violating template and summed in phase to extract the relevant signal power 
$\rho$. In other words, in the previous subsection we assumed a follow-up of each detected signal with  
a family of CS parity-violating, resonantly-amplified templates possessing two additional parameters 
($f_{p,s},f_{p,0}$) or equivalently $(\dot{\theta}_{0},\ddot{\theta}_{0})$.  In this section, we investigate constraints
on parity violation if all matched filtering is performed with parity-preserving, GR templates.

Given a GW detection, matched filtering with GR templates will return a set of best-fit 
parameters $\vec{{\lambda}}_{\GR}$ that maximize the SNR $\rho_{\rm{max}}$. In the absence of noise, the values of 
$(\dot{\theta}_{0},\ddot{\theta}_{0})$ that lead to $\tilde{\rho} =  \rho_{\rm{max}}$ are simply those on the diagonal of
Fig.~\ref{fig:amp}, i.e.,~$\ddot{\theta}_{0} = H_{0} \dot{\theta}_{0}$. In the presence of noise, however, random fluctuations in
the amplitude can resemble a resonant amplification. Therefore, one can search
for the largest values of $(\dot{\theta}_{0},\ddot{\theta}_{0})$ that lead to a $\tilde{\rho}$ that is consistent with 
$\rho_{\rm{max}}$. Such values of $(\dot{\theta}_{0},\ddot{\theta}_{0})$ then serve as an upper limit or constraint
on the magnitude of parity violation in the signal.

Such a study leaves out the fact that a GR matched filtering calculation unavoidably 
induces a fundamental error in the estimation of GR parameters, if the signal is parity violating. 
This is a manifestation of fundamental bias~\cite{Yunes:2009ke}. In other words, 
the true physical parameters of the signal $\vec{{\lambda}} = \vec{{\lambda}}_{\GR} + \delta\vec{{\lambda}}$, 
where $\delta\vec{{\lambda}}$ is the aforementioned error. We shall here ignore this error and
leave an investigation of fundamental bias to a future publication.

As in the previous subsection, given a source with $|\xi| = 1$ and a single detector, we cannot distinguish 
between a purely right or purely left circularly-polarized signal, and thus, the SNR $\tilde{\rho}$ we would detect is
\begin{eqnarray}
\tilde{\rho} &= & 4\int_{0}^\infty
df \frac{|\tilde{h}_{GR}|^2}{S_h(f)} \text{Re}\left[  \frac{\calF_{\RL}(\eta_s, k)}{\calF_{\RL}(\eta_{0},
    k)}\right]^{1/2} \equiv \hat{\rho} \; {\calAmpMatch_{\RL}}\,, 
\end{eqnarray}
where
\begin{eqnarray}
{\calAmpMatch_{\RL}} &\equiv& \frac{1}{I_7} \int_{0}^\infty df \frac{f^{-7/3}}{S_h(f)}\text{Re}\sqrt{
\frac{1-f/(a_{s} f_{p,s})}{1-f/(a_{0} f_{p,0})}} \,.
\label{tilde-G}
\end{eqnarray}
Notice that we have here crudely assumed the resonantly amplified waveform is projected onto a pure GR signal model
with identical parameters. To improve the realism of this calculation, one could maximize over time and phase of 
coalescence, but we shall not do this here.

Unlike the intrinsic SNR of the previous subsection, the integral defining $\tilde{\rho}$ requires no regularization. 
This is because ${\calAmpMatch_{\RL}}$ differs from $\calAmp_{\RL}$ in that the former contains a square root, 
while the latter contains an absolute value. This square root leads to a well-defined integral, as we explore in more
detail in the Appendix. All other details of the numerical implementation are the same as in the previous subsection.

When resonant frequencies are very large (i.e.,~in the linear regime, where constraints on $\theta'$ are  weak), 
then Eq.~\eqref{tilde-G} reduces to 
\begin{eqnarray}
\calAmpMatch \approx 1+ \frac{\lambda_{\RL}}{2} \left<v\right>,
\end{eqnarray}
which is precisely half the amplification seen using an optimal signal model in the previous subsection.
We have computed figures similar to Fig.~\ref{fig:amp} and we indeed find that the constraints quoted in the previous
subsection deteriorate at most by a factor of $2$. Of course, when one maximizes over extrinsic and intrinsic parameters,
the match will increase, and thus the constraints will improve. 

\section{Extension and discussion}
\label{Ext-disc}

In this section we discuss estimates of various factors that improve the order of magnitude
results from the previous section. We also discuss limiting noise sources that could potentially
deteriorate the constraints.

\subsection{Constraints from single short GRBs}

A network of $N_d$ GW detectors using coherent data analysis is more
sensitive than one detector by a factor $\sqrt{N_d}$ in $\rho$, neglecting
differences in orientation and noise spectrum.
The four-detector LIGO-Virgo network thus gains about a factor of 2 for a
typical sky position, though the full beam pattern is
complicated~\cite{Stuver:2006gr,CutlerFlanagan:1994}. Moreover, this network also is more robust than
a single detector if the GRB is viewed off-axis~\cite{Dalal:2006qt,
Seto:2005xb}.

A BH/NS merger takes advantage of the mass dependence of
the threshold~(\ref{single}), $\mathcal{M}^{5/6}$, which goes as the square
root of the total mass.
Thus a $30~M_\odot$ BH ($\mathcal{M} = 4.7~M_\odot$) would improve
the threshold further by a factor of about 4.
Above $30~M_\odot$ the improvement drops because the system starts merging
before the frequency has swept through the LIGO band and because
our analysis assumes no spin-orbit precession.  If the BH is rapidly spinning and that spin is misaligned with
the orbit, precession will lead to a time-dependent left/right
polarization ratio, reducing $\rho$ from the on-axis case
and complicating the analysis.  Spin-orbit misalignment could be large for dynamically-formed binaries.
In BH-NS binaries formed from isolated binary evolution, spin and orbital angular momenta should usually be 
aligned and precession should be relatively unimportant; see e.g.,~Fig.~6 in \cite{2004PhRvD..69j2002G}.

Observations and population arguments  suggest that GRB associated with NS/NS mergers have a narrow but nonzero range of
opening angles; population arguments suggest narrow jets
$\sigma_{\xi^2} = 0.02$~\cite{2006ApJ...650..261S}, while  the lack of jet breaks in
afterglows suggests that some jets may be as wide as $25^o$ \cite{2006ApJ...653..462G} (which is still $\xi\approx0.9$).
The finite collimation angle of the GRB will lead to a distribution of
$\xi^2$ which will degrade the threshold if $\sigma_{\xi^2} \approx 1/\rho$
since $\sigma_{\xi^2}$ adds roughly in quadrature with $\sigma_D/D$.  Additionally,  because 
 $\xi^2 < 1$, these unknown angles can only reduce $\rho$, not increase it, introducing a weak
 negative bias towards smaller amplitudes.   Without independent direct or a priori constraints on $\xi$, any short GRB
amplitude measurement by itself can only constrain the particular combination 
$\lambda_{\RL} \left<v\right> + (\lambda_{\RL} \xi - 1)$, as
\begin{eqnarray}
\frac{\rho}{\hat{\rho}(\xi)}&\approx& 1 + \lambda_{\RL} \left<v\right> 
\left[1 - \frac{1}{2} \frac{F_+^2 - F_\times^2}{F_+^2 + F_\times^2} (\lambda_{\RL} - \xi)^2
\right. 
\nonumber \\
&+& \left. O(\lambda_{\RL}-\xi)^3 \right]\,,
\end{eqnarray}
where we have explicitly expanded Eq.~\eqref{eq:amp:linear} for a nearly face on, right- or left-handed source
 $\xi \approx \lambda_{\RL}$ and used that $\hat{\rho}(\xi) \sim \hat{\rho}(\lambda_{\RL}) [1 + \lambda_{\RL} (\xi - \lambda_{\RL})]$. 
 In practice, however, the LIGO-Virgo
network (or any network of differently aligned detectors) is sensitive to both GW polarizations, and by
projecting directly onto a circularly-polarized basis, this network can directly constrain $1-\xi\simeq O(1/\rho)$.   
[A similar effect was observed in~\citet{Seto:2005xb} and~\citet{Dalal:2006qt}, where the LIGO-Virgo
network (or any network of differently aligned detectors) can reduce the
systematic errors in distance measurement to short GRBs induced by wide jets.]
The net effect of a narrow  distribution in $\xi$ is therefore simply to weaken any single-observation constraint on 
$\left<v\right>$ by roughly a factor of 2, from $1/\rho$ to $O(2)/\rho$.

Another source of error is due to mismodeling of the signal because of the neglect of higher harmonics present for example
in PN amplitude corrections~\cite{VanDenBroeck:2006qu}. Such corrections appear in the stationary-phase of the Fourier transform as modifications
to the amplitude of the form ${\cal{A}} \to {\cal{A}} \left( 1 + \zeta_{1/2} u^{1/3} + \zeta_{1} u^{1/3}  + \zeta_{3/2} u \right)$, where $u = 2 \pi M f$ 
is the reduced GW frequency and $\zeta_{n}$ are $n$th order PN amplitude coefficients, which depend on the mass ratio, the inclination angle 
and the polarization angles. Notice that the $1.5$PN correction, $\zeta_{3}$, is proportional to $f$, which immediately
suggests a strong-correlation and degeneracy with CS amplitude birefringence. This degeneracy should be affect the constraints quoted above much
because the mass ratio is primarily determined by the phase at $1$PN order, while $<v>$ does not enter the phase at all. 
A full study of such possible degenerate behavior would require a full Fisher or Bayesian analysis that we shall carry
out elsewhere.\optional{\footnote{\label{foot:Fisher:WhatWeShouldDo}The measurement accuracy of
    $\dot{\theta}_0,\ddot{\theta}_{0}$ should strictly be determined using a Fisher matrix, as in
    \citet{Alexander:2007kv}.  Further, this Fisher matrix should be constructed incorporating all terms comparable to
    and potentially correlated with the amplification discussed here, including pure GR post-Newtonian higher harmonics
    as well as
    modified-gravity wave-generation effects.  We will
    provide the relevant wave generation corrections and this matrix in a forthcoming paper. }}

The electromagnetic determination of the distance to the GW/GRB also has errors.
The latest augmented WMAP results~\cite{Jarosik:2010iu} imply $H_0 =
70.4\pm1.3$~km~s$^{-1}$~Mpc$^{-1}$ at 95\% confidence.  The intrinsic uncertainty in the Hubble constant translates to a relative
distance uncertainty of  roughly   $\delta H_{0} = 1.3/70.4\lesssim 1.8 \%$,
 far below the intrinsic SNR uncertainty $2/\rho$ for all but  extremely bright sources ($\rho\gtrsim108$).
Peculiar velocities of host galaxies, internal orbital velocities within galaxies and galaxy clusters,  
and kick velocities of binaries due to supernova kicks should all
be well under $1500{\rm km/s}\simeq 0.005 c$ \cite{2008NewAR..51..884M}, 
corresponding to a distance error of about 6\% in a Hubble redshift of $z=0.08$.
Only for extremely bright sources or very close double NS binaries will non-cosmological velocities produce a
noticeable error.  

A more significant and less well controlled distance error arises through the redshift determination process itself.
Currently short GRB distances can be estimated only through the redshift to electromagnetically associated hosts, and
therefore depends critically on  correct host identification
\cite{GalYam:2005jf,NakarReviewArticle2006,Gehrels-shortgrb-SwiftReview-Mid2007}.  For moderate to high redshift
sources, particularly where short GRBs are assumed ejected from their putative host galaxy, associations and therefore
redshift measurements could be significantly confused \cite{2009ApJ...705L.186Z}.
Fortunately, for the distances to which even advanced gravitational detectors
are sensitive, relatively few galaxies can conceivably be
associated with a coincident gravitational and electromagnetic signature (i.e.,
adopting a fiducial galaxy density $n_{gal}=0.01 \; \unit{Mpc}^{-3}$, the number of galaxies in the electromagnetic
sky-position angular extent $\theta$ and radial extent $D$  error cone is of order $n_{gal} D^3
\pi \theta^2/3 = {\cal{O}}(1)$ for $D=600\unit{Mpc}$ and $\theta\simeq 1\unit{arcmin}$).  We therefore assume any
future coincident short GRB and GW signal arising within $z<0.4$  
will have a correct host association and therefore a reliable redshift measurement.

Let us then summarize the enhancement and deterioration factors we have just discussed. A network of $N_{d}$
detector enhances the bound by a factor of $\sqrt{N_{d}}$, while a BH/NS event enhances the constraint by 
a factor of $4$. On the other hand, an unknown jet width deteriorates the constraint by a factor of $2$, while
a suboptimal search strategy also worsens the bound by a factor of $2$. Putting all of these together, we expect
a modification of the constraints quoted earlier by a factor of $1/\sqrt{N_{d}}$ for a NS/BH event or $4/\sqrt{N_{d}}$
for a NS/NS event (i.e., from $\left<v\right> < 1/\hat{\rho}$ to $1/( \hat{\rho} \sqrt{N_d})$ for a NS/BH event, where 
$\hat{\rho}$ is the single-detector SNR).

Putting all of this together, we arrive at the following constraints. 
A single detector with advanced LIGO sensitivity measuring a non-resonant, NS/NS inspiral signal 
seen directly overhead with unknown jet width leads to a $1\sigma$ constraint of 
$|\dot{\theta}_0-\ddot{\theta}_0/H_0| \leq 2400 \; \unit{km}$. A four-detector network would 
improve this limit to $|\dot{\theta}_0-\ddot{\theta}_0/H_0| \leq 1200 \; \unit{km}$. 
A BH-NS merger, seen at comparable sensitivity over a longer baseline, would reduce the limit  
to $|\dot{\theta}_0-\ddot{\theta}_0/H_0| \leq 300 \; \unit{km}$. All of this has assumed an optimal search strategy, but
with a suboptimal one, i.e.,~using pure-GR templates, one would only reach a bound of 
$|\dot{\theta}_0-\ddot{\theta}_0/H_0|<600\unit{km}$ for BH/NS events or $<2400 \; \unit{km}$ for NS/NS ones.

Since the maximum $(\dot{\theta}_0,\ddot{\theta}_0)$ allowed by these 
bounds would lead to a resonance in band, the absence of such a resonance could allow $(\dot{\theta}_0, \ddot{\theta}_0)$ to 
be constrained much better. In fact, this could constrain $|\dot{\theta}_0-\ddot{\theta}/H_0| \leq 50 \; \unit{km}$ to $1\sigma$ 
assuming an optimal search strategy (see e.g.,~Fig.~\ref{fig:amp} for $2\sigma$ constraints). If we additionally 
assume a four-detector coincident GW/GRB measurement of a NS/BH signal, the absence of a resonance could lead to a bound 
of\footnote{This constraint is an optimistic one, since if this bound is saturated, the resonance occurs near $2 \; {\rm{kHz}}$.
At such high frequencies our calculations is limited by systematic errors due to incorrect waveform modeling,
since one should in principle include the plunge and merger parts of the waveform in the modeling as well.} 
$|\dot{\theta}_0-\ddot{\theta}_0/H_0| \leq 25 \; \unit{km}$, even allowing for an unknown jet width and a suboptimal search strategy.

\subsection{Population constraints }

Combining $N_0$ GW/GRB observations will improve the CS threshold roughly
as $\sqrt{N_0}$.
Birefringence would manifest as a {\emph{bimodal distribution}} of
$\rho/\hat{\rho}$, with peaks on either side of unity.
If GRB jets are poorly collimated, both peaks of the distribution of
$\hat\rho/\rho$ would be smeared downward due to the distribution of $\xi$ (see above).

Advanced detectors will observe very many short GRBs with GW associations.  
Roughly  $R_{sky}\simeq 100-200$ short GRBs occur on the sky per year  that produce enough flux to be detected by the current
generation of satellites \cite{PSgrbs-popsyn,1999ApJS..122..465P,GuettaPiran2005}.  
As a first approximation, host associations suggest that the
\emph{satellite-detected} short GRB population is roughly linearly distributed in redshift out to $z\simeq O(1)$
\cite{GuettaPiran2006,GuettaPiran2005,PSgrbs-popsyn,NakarReviewArticle2006}.   Therefore, adopting an isotropic network
horizon\footnote{Because each short GRB is presumed optimally oriented, the range to which it can be detected is
significantly greater than the angle-averaged range to which a typical source could be observed.  In practice, the
distance to which short GRBs could be detected will be greater still, because when a shorter stretch of data is searched, 
the detection threshold can be set significantly lower~\cite{Abbott:2005yy,Abbott:2008zzb}.}  distance $D_{gw}>200{\rm Mpc}$,
large enough to enclose many binary mergers per year,  roughly 
\begin{eqnarray}
R_{gw+GRB} &\simeq& \frac{f_{sky}}{z_{max}} D_{gw}H_0 R_{sky}, \\
 &\simeq&  \frac{20}{{\rm yr}} \frac{f_{sky}}{z_{max}} \frac{D_{gw}}{445 \; {\rm Mpc}} \frac{R}{200 \; {\rm yr}^{-1}} \nonumber
\end{eqnarray}
satellite-detected  short GRBs could be detected by a GW interferometer network.   
Even under more conservative assumptions including the local event rate and expected beaming angles, 
a significant number of NS-NS or BH-NS binaries should be coincident with
a short GRB event: more than  $>O(10\%)$ of all GW detections.  Given the latest event rate predictions \cite{O'Shaughnessy:2006wh,PSellipticals}, at least 2 should be detected per year. To summarize, a population of at least $4$ short GRBs 
could improve constraints on $\dot{\theta}_0$ by a factor $\sqrt{4}= 2$.

Combining this factor of two and those of the previous subsection, we find that parity violation could be constrained 
to the level of $\left<v\right> < 1/(\hat{\rho} \sqrt{N_d N_{0}})$ for a NS/BH event. Modeling parity violation with CS gravity and
first adopting a non-resonant parity-violating event, 
it is reasonable to expect that after two calendar years (the length of the LIGO S5 run) an advanced LIGO-Virgo network
operating in coincidence with GRB observatories would yield a 1$\sigma$ confidence detection of or upper limit on 
$|\dot{\theta}_0-\ddot{\theta}_o/H_0|$ of $150 \; \unit{km}$ for BH-NS ($600 \; \unit{km}$ for NS-NS) using optimally matched 
templates, or $300 \; \unit{km}$ for BH-NS ($1200 \;  \unit{km}$ for NS-NS) using a conventional pure-GR data analysis strategy.
Again, as the maximum $(\dot{\theta}_0,\ddot{\theta}_0)$ allowed by the aforementioned constraints would lead to a resonance 
in band, the absence of resonances could allow a much better constraint. In the neighborhood of $\ddot{\theta}_{0} = 0$, 
the absence of such resonance could yield a bound on $\dot{\theta}_{0}$ of ${\cal{O}}(10 \; \unit{km})$. 

The projected constraints on CS gravity can be thus more than an order of magnitude better than current Solar System bounds
placed with the LAGEOS satellites~\cite{Smith:2007jm}. Although none of these constraints are competitive with 
the one recently placed with the double binary pulsar~\cite{Yunes:2008ua}, the latter only samples parity violation
in the neighborhood of this binary, while the Solar System one is sensitive to parity violation only
in the vicinity of the Earth-Moon system. The test proposed here would allow for constraints of the dynamical sector
of parity violating theories along the GW geodesic, thus exploring a much larger region of the Universe. 

\subsection{Temporal-Spatial Variability of Parity Violation}

In this paper, we have proposed limits on $\dot\theta_0$ and $\ddot\theta_0$ assuming (i) that $\dot{\theta}$ varies on cosmological timescales and (ii)
homogeneously and isotropically throughout the Universe. In principle, neither assumption need apply, as the $\theta$ field that parametrizes
parity violation in CS gravity should be prescribed by the more fundamental theory of which CS gravity is only an effective model. 
If the $\theta$ field does not obey the symmetries of the background (homogeneity and isotropy in the cosmological case), then the background
metrics employed, on which GW propagate, need not be solutions to the effective theory. More general $\theta$ fields would possibly require a
re-analysis of the appropriate cosmological solutions and the propagation of GWs in them. 

With this caveat in mind, the more generic expressions of Eq. (\ref{generalv}) that parameterize generic parity violating effects via the amplification
process of Sec.~\ref{res-mag} suggest that the proposed constraints are primarily on a ratio of $\calF_{\RL}$, with one
factor at the source (the numerator) and another, more critical one at the detector (the denominator).  These
two factors can be separated.  For example,  the resonant frequency and resonant bandwidth in our detectors are set primarily by the present-day evolution
of $\theta$.    If a resonance is indeed in band, \emph{every} source with some amount of resonant polarization will
be amplified roughly the same way; stacking power from multiple detections could constrain the presence of a universal
spectral feature and potentially directly constrain $\dot{\theta}_0$, independent of its cosmological behavior, to lie
outside the wavelength region to which LIGO is sensitive.
On the other hand, the difference between the values of $\dot{\theta}$ at present and the source determine how close a zero
in the numerator is to the resonance and  thus the amplification ${\calAmp}_{\RL}$; cf. Eq. (\ref{eq:Amp:Approx}).  Therefore, if the
detected SNRs do exhibit some  amplifications $\rho/\hat{\rho}$ versus $D$ that are \emph{not} roughly uniformly
increasing with distance, these measurements can be fit to either a functional form (if the points trace a curve) or a
distribution (if the points have excess scatter) for $\dot{\theta}$.  As a concrete example, so long as resonances are at
frequencies far higher than in band
($\kappa\theta'\ll 1$) and the source is at low enough redshift that cosmological terms can be ignored ($a=a_s\approx
1$), the relative amplification  in SNR due to CS birefringence is [cf. Eq. (\ref{eq:v:Simple2})]
\begin{eqnarray}
\rho/\hat{\rho}&\approx& 1  + \lamRL \pi \left<f\right> (\theta'_0 -\theta'_s),
\end{eqnarray}
which applies to any $\theta'(\eta)$ which varies on scales faster than cosmological timescales.  
This would then allow for generic tests of gravitational parity violation, irrespective of the specific theoretical model.

\optional{\subsection{Tests without EM distances or circular polarization}
The test described above is straightforward precisely because it relies on special input:
electromagnetically-determined distances to sources assumed to emit circularly polarized gravitational waves.  
More sophisticated data analysis strategies can relax both conditions.  For example, even without a known distance, the
population of nearly circularly polarized sources with known sky location naturally yields a bimodal distribution of
SNR, with a bias proportional to a (distance-independent) $\delta \rho$; the distribution of left and right handed
sources therefore always constrains $\delta \rho$ and therefore birefringence.  Likewise, all binary sources have a
mixture of left- and right-handed signal power; birefringence implies a network will recover more of one handedness than
another, overall.  The relative proportion of left- versus right-handed power therefore provides another test.}

\acknowledgments

We thank D. Fox and M. C. Miller for helpful discussions and N. Cornish for
helpful comments on the manuscript.
This work was supported by NSF grants PHY-0555628, PHY-0748819,  
and PHY-0855589, and by the Center for Gravitational Wave Physics under NSF
cooperative agreement PHY-0114375. N.Y. was supported by National 
Science Foundation award PHY-0745779. R.O. was supported by National Science Foundation 
award PHY 0653462 and the Center for Gravitational Wave Physics.
S.A was supported by NSF Grant PHY-0901925. 
B.J.O. was also supported by the LIGO Visitors Program. LIGO was
constructed by the California Institute of Technology and Massachusetts
Institute of Technology with funding from the National Science
Foundation and operates under cooperative agreement PHY-0107417. This
paper has been assigned LIGO Document Number P1000051.

\appendix*
\section{Singular Behavior and Resonances in CS Gravitational Waves}
\label{app}

Parity violating GWs need not present a pole in the frequency domain, but the theory we use to exemplify
such violation, CS gravity, does. As shown in Sec.~\ref{CS-basics}, 
such singularities arise due to poles in the modified field equations, i.e.,~Eqs.~\eqref{wave2} and~\eqref{wave}. 
The effect of such singularities was analyzed in Sec.~\ref{res-mag} where it was seen to lead
to a resonant magnification or amplification of the SNR. That section, however, does not explain why 
such amplification is finite; this is the topic of this appendix. We begin with a basic
introduction to resonances and the non-uniform validity of perturbation theory, as this topic may be
unfamiliar to the GW community, following mostly~\cite{Bender}. 
We then proceed to the study of singularities in CS gravity, their characterization and treatment. 

\subsection{Resonances and the Nonuniform Validity of Perturbative Expansions }

Consider the differential equation for a driven harmonic oscillator
\begin{equation}
\ddot{y} + y = \cos{\omega t}\,,
\label{DHO}
\end{equation}
whose solution is
\begin{equation}
y(t) = A \cos{t} + B \sin{t} + \frac{\cos{\omega t}}{1 - \omega^{2}}.
\end{equation}
This solution is valid only when $|\omega| \neq 1$, as in the limit $\omega \to 1$, the
oscillator absorbs large quantities of energy and the amplitude of the oscillations grow unbounded. 
The solution to Eq.~\eqref{DHO} at resonance ($|\omega|=1$) is
\begin{equation}
y = A \cos{t} + B \sin{t} + \frac{t}{2} \sin{t}\, \qquad |\omega|=1. 
\end{equation}
Notice that the divergent term has been replaced by a regular, {\emph{secular}} contribution. Such a
secular term grows with time and is unbounded in the limit $t \to \infty$, as then the oscillator
absorbs energy without limit. 

The emergence of secular terms in the solution to certain differential equations signals the breaking
of uniform perturbation theory. In order to exemplify this concept in more detail, consider Duffing's
differential equation
\begin{equation}
\ddot{y} + y + \epsilon \; y^{3} = 0\,,
\label{DuffingEq}
\end{equation} 
with $\epsilon \ll 1$ and the boundary conditions $[y(0),\dot{y}(0)] = [1,0]$. Although the $\epsilon \; y^{3}$ 
term cannot be interpreted as a driving force with a certain natural frequency, we shall show next that its
perturbative solution unavoidably introduces secular growth, just as in the case of the driven harmonic
oscillator at resonance.

Let us assume the following perturbative ansatz for the solution to Duffing's equation
\begin{equation}
y(t) = \sum_{n=0}^{\infty} \epsilon^{n} y_{n}(t).
\end{equation}
Order by order in multiple scale analysis, Duffing's equation becomes $\ddot{y}_{0} + y_{0} = 0$ to ${\cal{O}}(\epsilon^{0})$, 
and $\ddot{y}_{1} + y_{1} = -y_{0}^{3}$ to ${\cal{O}}(\epsilon)$. The
solution to the zeroth-order equation that satisfies the above initial conditions is simply $y_{0}(t) = \cos{t}$. 
The solution to the first-order equation that satisfies the same initial conditions leads to the full solution
\begin{equation}
y(t) = \cos{t} + \left[ \frac{1}{32} \left( \cos{3 t} - \cos{t} \right) - \frac{3 \, t}{8} \sin{t} \right] \epsilon + {\cal{O}}(\epsilon^{2}).
\label{1stOsol}
\end{equation}
Notice that a secular term has appeared in the first-order solution, i.e.,~the last term inside the square-brackets 
is unbounded as $t \to \infty$. This solution is valid to ${\cal{O}}(\epsilon^{2})$ {\emph{for fixed $t$}}, 
but as one considers times $t = {\cal{O}}(1/\epsilon)$ or larger, Eq.~\eqref{1stOsol} ceases to be valid.

The non-uniform validity of the perturbative solution leads to an inaccurate approximation to the exact solution
at late-times, as the exact solution is actually bounded for all times.  To show this, one can construct an {\emph{energy integral}}
and show that it is bounded. In the case of Duffing's equation, one can multiply Eq.~\eqref{DuffingEq} by $\dot{y}$ and then rewrite
it as
\begin{equation}
\frac{1}{2} \dot{y}^{2} + \frac{1}{2} y^{2} + \frac{1}{4} \epsilon y^{4} = C,
\label{boundedness}
\end{equation}
after one time integration with constant $C$. Equation~\eqref{boundedness} can be thought of as a closed bounded orbit in the phase 
space with coordinates $(y,\dot{y})$. Due to the initial conditions, one easily finds that $C = (2 + \epsilon)/4$. 
Since each term on the left-hand side is positive if $\epsilon > 0$, then $y^{2} < 2 C$ and $y$ is bounded 
by $y < (1 + \epsilon/2)^{1/2}$. We have thus proved that although the perturbative solution presents divergent behavior at late times
due to the secular term arising at resonance, the exact solution is properly bounded. Another way to show this is to partially sum all
the divergent, secular terms in the perturbative solution. In the case of Duffing's equation, one would find that these terms sum up to
$\cos[t(1 + 3 \epsilon/8)]$, which is properly bounded at $t \to \infty$. 

This subsection has taught us several important lessons. 
Resonant solutions to differential equations can be studied close to their singularities via perturbation theory. 
Such a study might reveal secular growth in the perturbative solutions, but this does not mean that the solution, or the 
theory from which this derives, is unstable. Instead, such unbounded growth is an indication that the perturbative solution has
a non-uniform region of validity. The construction of positive energy integrals allows us to prove that solutions are bounded from above,
and that secular growth is artificial.
 
\subsection{Boundedness of CS Resonances}

Let us now consider the resonant structure of the CS differential equations for wave propagation and their perturbative solution.
In Sec.~\ref{CS-basics} we found that the differential equations for CS phase evolution have a singular driving force, i.e.,~the right-hand
sides of Eqs.~\eqref{wave2} and~\eqref{wave} diverge in the limit $\lambda_{\RL} k \dot{\theta} \to 1$. Later in Sec.~\ref{gen-amp-bire} 
we found that the exact solution to these differential equations are also singular in this limit, i.e.,~Eq.~\eqref{exactsol} diverges as 
${\cal{S}}_{\RL} \to 0$. 

The perturbative, approximate solutions to the singular CS phase evolution equations are in fact regular, 
although they lead to secular growth. This can be seen from Eqs.~\eqref{deltaphiflat} and~\eqref{deltaphi}, which
are proportional to the GW travel time, or equivalently distance, so that as $t \to \infty$, the perturbative solution is unbounded. 
Such secular growth is nothing but an indication of the non-uniform validity of the perturbative solution. In fact, from the structure of
Eqs.~\eqref{deltaphiflat} and~\eqref{Tay-Exp-theta}, we can deduce that the approximation seizes to be valid when the GW has travelled 
a distance $D = {\cal{O}}(f_{0}^{-1}\ddot\theta^{-1}_0)$ or $D = {\cal{O}}(f_{0}^{-1} H_{0}^{-1} \dot\theta^{-1}_0)$.

Let us now try to show boundedness of the solution through the construction of an energy integral. 
Equation~\eqref{waveh3} can be rewritten as
\begin{equation}
h''_{\kappa} + 2 \frac{d}{d\eta} \left(\ln {\cal{S}}_{\RL}\right) h'_{\kappa} + \kappa^{2} h_{\kappa} = 0\,.
\end{equation}
For simplicity, we take $h$ to be real and multiply the above equation by $h'$. Integrating once we find
\begin{equation}
(h'_{\kappa})^{2} + 4 \int (h'_{\kappa})^{2} \left[\frac{d}{d\eta} \left(\ln {\cal{S}}_{\RL}\right)\right]  d\eta
+ \kappa^{2}  h^{2}_{\kappa} = 2 C\,.
\end{equation}
Choosing initial conditions $h_{\kappa}(0) = 1$ and $h'_{\kappa} = 0$, we find that $C = \kappa^{2}/2 \geq 0$.
It then follows that the solution is bounded from above by
\begin{equation}
h_{\kappa} < \sqrt{1 - \frac{4}{\kappa^{2}} \int (h'_{\kappa})^{2} \left|\frac{d}{d\eta} \left(\ln {\cal{S}}_{\RL}\right)\right|  d\eta }
\end{equation}
provided the modified scale factor is positive:
\begin{equation}
\frac{{\cal{S}}_{\RL}'}{{\cal{S}}_{\RL}} = \frac{\dot{\cal{S}}_{\RL}}{{\cal{S}}_{\RL}} 
= \frac{1}{2} \frac{2 H - \lambda_{\RL} k \left(\ddot{\theta} + H \dot{\theta}\right) }{1 - \lambda_{\RL} k \dot{\theta}} > 0\,.
\end{equation} 

The positivity of the effective Hubble parameter $\dot{\cal{S}}_{\RL}/{\cal{S}}_{\RL}$ depends on the size of $\dot{\theta}$. 
This quantity is indeed positive when $\dot{\theta} = 0 = \ddot{\theta}$ and the universe is expanding such that $H > 0$. Moreover, this quantity is also
positive away from the resonance, i.e.,~if $\dot{\theta} \ll \lambda_{\RL}/k$, as then the CS modification is a small deformation of the GR result. 
At resonance, ($\dot{\theta} = \lambda_{\RL}/k$, $\ddot{\theta}=0$), the positivity of $\dot{{\cal{S}}}_{\RL}/{\cal{S}}_{\RL}$ 
depends on whether $\dot{\theta}$ approaches $\lambda_{\RL}/k$ from below or above. If approached from below, 
$\dot{\theta} = \lambda_{\RL} (1 - |\epsilon|)/k$ as $\epsilon \to 0_{+}$, then $\dot{{\cal{S}}}_{\RL}/{\cal{S}}_{\RL} > 0$, while the opposite is
true if approached from above, $\dot{\theta} = \lambda_{\RL} (1 + |\epsilon|)/k$ as $\epsilon \to 0_{+}$. 

For strong CS parity violation, i.e.,~$\dot{\theta}> {\cal{O}}(1/k)$, $\dot{{\cal{S}}}_{\RL}/{\cal{S}}_{\RL} > 0$ for all $\dot{\theta} \geq 2 \lambda_{\RL}/k$, 
but we cannot prove the solution is bounded via this energy integral method if $\lambda_{\RL} k \dot{\theta}$ is in the interval $[1,2]$. A full treatment of
strong CS parity violation, however, requires that one includes higher-order $\theta$-interactions in the action, as Eq.~\eqref{CS-action} is an 
{\emph{effective}} theory valid in the weak coupling limit only. Such interactions will contribute to the energy integral, particularly in the strong coupling
limit. Since we do not have a higher-order completion of CS gravity, we cannot prove boundedness in this limit.

In spite of the apparent secular growth of the perturbative solution in the weak CS coupling limit, 
we have here shown that the exact solution remains bounded in this limit. The secular growth observed 
is then nothing but an artifact of the perturbative expansion and should not be seen as an indication of 
an instability in the theory. 

\subsection{Trans-Singular CS Resonances}

Proving that the exact solution to the CS phase evolution equation remains bounded does not tell us how to deal with its singularity.
However, as in the case of the driven harmonic oscillator of Eq.~\eqref{DHO}, the solution we found away from resonance is not valid
at resonance. In order to find the solution close to the singularity, we must solve the phase evolution equation perturbatively, restricting
attention to a neighborhood close to the pole. 

The general form of a parity violating wave equation  
\begin{equation}
h''_{\RL,\kappa} + \kappa^2 h_{\RL,\kappa} + \frac{\calF_{\RL}'}{\calF_{\RL} } h'_{\RL,\kappa} = 0
\label{mainwaveeq}
\end{equation}
has a pole when $\calF_{\RL}$ vanishes. 
Equation~\eqref{mainwaveeq} is nothing but Eq.~\eqref{waveh} with the replacement 
$\mathcal{H} \to \mathcal{S}_{\RL}'/\mathcal{S}_{\RL} \equiv  \calF_{\RL}'/(2 \calF_{\RL})$, 
where we recall that the squared amplitude $\calF_{\RL} \equiv \mathcal{S}^{2}_{\RL}$.
With this notation, recall that in CS gravity $\calF_{\RL} \equiv a^2 - \lamRL \kappa \, \theta'$.
Let us linearize the source function $\calF_{\RL}$ about the ($\kappa$-dependent) conformal pole 
$\eta_p(\kappa)$, defined via $a^2(\eta_{p}) - \kappa \, \theta'(\eta_{p}) = 0$, though the Taylor expansion 
$\calF_{\RL} = \calF_{\RL}|_{p} + \calF_{\RL}'|_{p} (\eta - \eta_{p})$, where the vertical bar stands for
evaluation at the pole. Equation~\eqref{mainwaveeq}  simplifies to a Bessel equation
\begin{equation}
\label{eq:ap:Bessel}
h''_{\RL,\kappa} + (\eta - \eta_p)^{-1} h'_{\kappa} + \kappa^2 h_{\RL,\kappa} = 0.
\end{equation}
Notice that by linearizing about $\eta = \eta_{p}$ we have lost the distinction between left and right-polarized waves,
as the $\lambda_{\RL}$ dependence in the source term appears only at second order in the Taylor expansion.
 
Solutions to this wave equation linearized about the pole can be straightforwardly obtained  
as a superposition of Hankel functions. Imposing a no-incoming-radiation boundary condition
at infinity, the solution reduces to the outward propagating, cylindrical wave solution
\begin{eqnarray}
h_{\RL,\kappa}(\eta) &=& {\cal C} \, H_0^{(1)}\left[\kappa \left(\eta-\eta_p\right)\right]\,,
\label{hankel-sol}
\end{eqnarray}
where ${\cal{C}}$ is a normalization constant. The quantity $H_0^{(1)}$ is a Hankel function of the first kind, 
not to be confused with the Hubble parameter at present time. Notice that this solution is discontinuous and singular at
$\eta=\eta_p$, a conformal time often reached after propagation on cosmological scales.  
Guided by analytic continuation, by conserved current across the singularity  (the Wronskian of
Eq. (\ref{eq:ap:Bessel})), and by the need for a nearly-Minkowski gravitational propagator in the neighborhood of
$\eta\approx \eta_p(f)$, we adopt this particular solution on both sides of the resonance.

One can reinterpret the solution in terms of propagators. In the neighborhood of some $\eta_s$, e.g.,~near the source, 
each spatial Fourier component is evolved forward with a simple scalar propagator:
\begin{eqnarray}
h_{\RL,\kappa}(\eta) &=& K_{\kappa}(\eta,\eta_s) \, h_{\RL,\kappa}(\eta_s), \\
K_{\kappa}(\eta,\eta_s) &\equiv& \frac{H_0^{(1)}\left[\kappa\left(\eta-\eta_p\right)\right]}{H_0^{(1)}\left[\kappa\left(\eta_s-\eta_p\right)\right]}.
\label{res-resolve}
\end{eqnarray}
The functional form of this equation resembles the solution to the differential equations of para-axial optics, 
where as initial data we require only the initial state $h_{\RL,\kappa}(\eta_{s})$. 
In fact, except for a fraction of a wavelength in the neighborhood of the pole $\eta \approx \eta_{p}$, 
the propagator can be well-approximated at late times by 
\begin{equation}
K_{\kappa}(\eta,\eta_s) \approx \sqrt{\frac{\eta_s-\eta_p}{\eta-\eta_p}}  e^{-i\kappa(\eta-\eta_s)}, \quad \kappa (\eta - \eta_{p}) \gg 1\,, 
\label{prop-exp}
\end{equation}
which can be obtained by either expanding the Hankel functions in $\kappa(\eta - \eta_{p}) \gg 1$ or solving Eq. (\ref{eq:ap:Bessel})
in this limit. 

The functional form of the propagator in terms of the source function $\calF_{\RL}$, namely
\begin{equation}
K_{\kappa}(\eta,\eta_s)\approx \sqrt{\frac{\calF_{\RL}(\eta_s)}{\calF_{\RL}(\eta)}} e^{-i\kappa(\eta-\eta_s)}\,
\label{prop-gen}
\end{equation}
was already derived in Sec.~\ref{gen-amp-bire}. Taylor expanding the source function about resonance 
\begin{equation}
{\cal{F}}_{\RL}(t)  \sim 1 - \lambda_{\RL} k_{p} \dot\theta - \lambda_{\RL} k_{p} \ddot{\theta} (t-t_{p}) = - \lambda_{\RL} k \ddot{\theta} (t-t_{p})\,, 
\end{equation}
where in the second equality we have used the resonance condition $\dot{\theta} = \lambda_{\RL}/k$,
we find that Eq.~\eqref{prop-gen} reduces exactly to Eq.~\eqref{prop-exp}. Notice however that Eq.~\eqref{prop-gen}
is formally valid away from the pole, while the true representation of the solution close to the pole is given by Eq.~\eqref{res-resolve}.

All of these solutions diverge when $\eta = \eta_p$, but the width of the resonance is extremely narrow in frequency space.  
This width can be approximated as $\delta f_{p} \simeq \dot{f}_{p} \delta t_{p}$, which for a minimal-width wave packet with 
$\delta t_{p} \delta f_{p} \simeq 1$,  implies $\delta f_p = {\cal{O}}(\dot{f}_{p}^{1/2})$.  Using that $f_{p} = 1/(2 \pi \dot{\theta})$ 
and $\dot{f}_{p} = -\ddot{\theta}/(2 \pi \dot{\theta}^{2})$, we find that $\delta f_{p} \approx {\cal{O}}[(H/\dot{\theta})^{1/2}]$, 
where we have assumed $\dot{\theta}$ evolves on cosmological scales $\ddot{\theta} = {\cal{O}}(H \dot{\theta})$. 
Evaluating these expressions today on Earth, we find that $\delta f_{p} = {\cal{O}}(10^{-8} {\rm Hz})$, where for $\dot\theta$ 
we have saturated the Solar System constraint $\dot{\theta} < 2000 \; {\rm km}$. 

Although any \emph{single, large-scale-coherent Fourier mode} is formally divergent,
any finite-duration, band-limited waveform always remains finite. This is because to obtain any finite-duration waveform
one must undo the Fourier decompositions performed in this paper [see e.g.,~Eq.~\eqref{wave-ansatz} or Eq.~\eqref{wave-ansatz2}]. 
The reconstruction of the full waveform then requires the integration of $h_{\kappa}(\eta)$ over all wave-numbers:
\begin{eqnarray}
h_{\RL}(\eta,\chi^{i}) &\equiv& \frac{1}{(2 \pi)^{3/2}} \int d\kappa \; h_{\RL,\kappa}(\eta) e^{-i \kappa_{i} \chi^{i}}\,,
\\ \nonumber 
&=& \frac{{\cal{D}}}{(2 \pi)^{3/2}} \int d\kappa \; 
K_\kappa(\eta,\eta_s)
 \kappa^{-7/6} e^{-i \kappa (\chi - \eta_{s})}\,,
\end{eqnarray}
where ${\cal{D}}$ is a new constant related to ${\cal{A}}$ in Eq.~\eqref{A-const}, which depends on the chirp mass, inclination angle and distance to the source,
and where we have assumed the wave propagates in the $\chi$ direction with conformal wavenumber $\kappa$. 
This integral can be split into three integrals: two pieces that avoid the singularity of the propagator  close to the pole
and one piece that integrates through this singularity.  Clearly, the first two pieces are finite, while the last piece
can be approximated using Eq. \eqref{res-resolve} in the neighborhood of the pole:
\begin{eqnarray}
h_{p,\RL} &\sim& \frac{{\cal D}}{(2 \pi)^{3/2}} \int_{\kappa_{p}^{-}}^{\kappa_{p}^{+}}  
d\kappa  \, 
 \frac{H_0^{(1)}\left[\kappa\left(\eta_{0}-\eta_p\right)\right]}{H_0^{(1)}\left[\kappa\left(\eta_s-\eta_p\right)\right]}
\frac{e^{-i \kappa (\chi - \eta_{s})}}{\kappa^{7/6}}\,,
\label{hp-1}
\nonumber \\
\end{eqnarray}
where we have defined $\kappa_{p}^{\mp} = \kappa_{p} \mp 2 \pi \epsilon$.
An asymptotic expansion of the Hankel function about zero argument reveals that
\begin{equation}
H_{0}^{(1)}(x) \sim \text{sign}(x)+  \frac{2i}{\pi} \left[\ln\left(\frac{|x|}{2}\right) + \gamma_{\rm E}\right], \qquad x \ll 1\,
\label{Hankel-asy}
\end{equation}
where $\gamma_{\rm E}=0.5772157\ldots$ is the Euler constant. Using this expansion, assuming the light-cone condition $\chi = \eta_{s}$, 
and pulling out of the integral all quantities that remain roughly constant near the pole, the integral of Eq.~\eqref{hp-1} becomes
\begin{eqnarray}
\label{final-int}
h_{p,\RL} &\sim& \frac{{\cal{E}}}{\kappa_{p}}  \int_{\kappa_{p}^{-}}^{\kappa_{p}^{+}} 
d\kappa \;  \left\{ \text{sign}(\kappa_p(\eta_0-\eta_p))
  \nonumber \right. \\ 
 &+& \left. \frac{2i}{\pi} \left[\ln\left(\frac{|\kappa_{p}\left(\eta_{0}-\eta_p\right)|}{2}\right) + \gamma_{\rm E}\right] \right\}
\nonumber \\
{\cal{E}} &\equiv& \frac{{\cal D}}{(2 \pi)^{3/2}} \frac{\kappa_{p}^{-1/6}}{H_0^{(1)}\left[\kappa\left(\eta_s-\eta_p\right)\right]}\,.
\end{eqnarray}
Using $\eta_{0} = \eta_{p} + {\cal{R}} (\kappa - \kappa_{p})$, we find that the most divergent term of the above integral is
proportional to 
\begin{equation}
\label{eq:hres:FiniteInTime}
h_{p,\RL} \propto {\cal{E}}  \; \frac{\epsilon}{\kappa_{p}}  \ln\left( \kappa_{p} {\cal{R}} \epsilon\right)
= {\cal{E}}  \; \frac{\epsilon}{\kappa_{p}}  \ln\left( \frac{\kappa_{p} \epsilon}{\delta \kappa_{p}^{2}} \right)
\,,
\end{equation}
where in the second relation we have used that ${\cal{R}} = {\cal{O}}(\delta \kappa_{p}^{-2})$.
We have then proved that any finite-duration, band-limited waveform always remains finite, as $h_{p,\RL}$ remains
finite when $\epsilon \to 0$.  Furthermore, the contribution of the inverse Fourier transform near the pole is 
clearly subdominant: to use a numerical example, saturating the Solar System constraint $\dot{\theta}_{0} = 2000 \; {\rm{km}}$, such that 
$\kappa_{p} \sim 150 \; {\rm{Hz}}$, and using a buffer of size $\epsilon = 10^{-8} {\rm{Hz}}$ around the pole, then $h_{p} \leq 10^{-8} {\cal{E}}$.

One might worry that although the pole does not affect the waveform itself, 
it might affect the calculation of the SNR, as this depends on the
integral of the square of the Fourier transform of the waveform.     
The \emph{approximate} temporal-only Fourier
transform of the propagator of Eq.~\eqref{res-resolve} in the stationary phase approximation is
\begin{equation}
\tilde{K}_\kappa(\omega|\eta_0) \equiv 
  \int d\eta K_\kappa(\eta,\eta_0) e^{i \omega \eta}  
 \approx  \sqrt{\frac{{\cal F}(\eta_0)}{{\cal F}(\eta_s)}}\delta(\omega-\kappa)e^{i\omega\eta_0}\,,
\end{equation}
where $\omega$ is conformal frequency. Then, the present-day temporal-only Fourier transform
$\tilde{h}(\omega)$ can be expressed in terms of the spatial Fourier transform $h_{\kappa}$ 
at the fiducial source time $\eta_0$:
\begin{eqnarray}
\tilde{h}(\omega)&=& \int \frac{d\kappa}{2\pi} \tilde{K}_\kappa(\omega) \; h_\kappa(\eta_0) \; e^{-i\kappa \chi}\,,
\end{eqnarray}
where $\eta_0-\eta_s=\chi$.  Except for an extremely narrow neighborhood near $\eta_0 \ne \eta_p(f)$, 
the stationary phase propagator and thus this model for $\tilde{h}(\omega)$ is an excellent approximation.
Substituting this expression into  Eq.~\eqref{srd} for SNR gives the
expression for resonant SNR adopted in the text [Eq.~\eqref{newResSNR}].  

As the denominator in the SNR inevitably includes a pole
at the zero of ${\cal F}_\kappa$ ($\kappa=\lambda_{\RL}/\theta'$),
this approximate integral indeed formally diverges logarithmically.
However, this logarithmic divergence arises by applying an approximate propagator and its Fourier transform
outside of their domain of applicability.  
In the neighborhood of $f_{p,s}$, the relevant propagator is approximately Eq.~\eqref{res-resolve}, modulo
higher-derivative corrections to ${\cal F}'/{\cal F}$ [cf. Eqs. \eqref{mainwaveeq}, \eqref{eq:ap:Bessel}].  

The temporal-only Fourier transform of the resonant propagator is proportional to the Fourier transform of the standard
Hankel function, keeping in mind this Fourier transform (unlike many used in classical scattering problems) is over a
two-sided argument:
\begin{eqnarray}
\tilde{K}(\omega)&=& \int_{-\infty}^{\infty} e^{i \omega \eta} \; \frac{H_0^{(1)}[\kappa(\eta-\eta_p)]}{H_0^{(1)}[\kappa(\eta_s-\eta_p)]} d\eta\,,
 \nonumber \\
\label{eq:prop:Fourier}
&=& \frac{e^{i\omega \eta_0}}{\kappa} \frac{\tilde{H}_0^{(1)}(\omega/\kappa)}{H_0^{(1)}(\kappa(\eta_s-\eta_p))}\,, \\
\tilde{H}_0^{(1)}(Y)&\equiv& \int_{-\infty}^{\infty} dX H_0^{(1)}(X) e^{i X Y}
\end{eqnarray}
The numerator, the Fourier transform $\tilde{H}_0^{(1)}$, is well-behaved and unaffected by the proximity of $\omega$ or
$\kappa$ to resonance ($\eta_0=\eta_p(\kappa)$).  The resonance condition enters only in the 
denominator. As  $H_0^{(1)}(X)$ has no zeros for real $X$, however, the time-domain Fourier transform of the
resonant propagator $\tilde{K}$ and the transform of the amplified waveform $\tilde{h}(\omega)$ have no poles 
on the real axis. This shows that even if the calculation where done exactly, the true SNR would remain finite. 
Moreover, given the extremely slow rate at which the resonant frequency evolves, 
we can approximate $\tilde{H}_0^{(1)}(\omega/\kappa)$ as proportional to
$\delta(\omega-\kappa)$. This implies that to a good approximation the Fourier 
transform of the \emph{time dependent} propagator is $K_\kappa(\eta_0,\eta_s) \delta(\omega-\kappa)$. 

With this reasonable approximations at hand, the SNR becomes
\begin{eqnarray}
\rho^{2}_{\RL} &=& 4 \int_{0}^{\infty} \frac{|\tilde{h}_{\RL}|^{2}}{S_{h}(f)} df
= 4 \int_{0}^{\infty} \frac{|{h}_{\RL,\kappa}|^{2}}{S_{h}(\kappa)} d\kappa\,,
\nonumber \\
&=& {\cal{D}}^{2} \int_{0}^{\infty} 
 |K_\kappa(\eta_0,\eta_s)|^2
 \kappa^{-7/3} \frac{d\kappa}{S_{h}(\kappa)}.
\end{eqnarray}
As before, let us split this integral into three pieces: two of them that avoid the singularity and one that integrates through it. 
In the neighborhood of the singular point, we once more approximate the propagator using Eq. \eqref{res-resolve}, so that
\begin{eqnarray}
\rho^{2}_{\RL,p} &\simeq& 4 {\cal{D}}^{2} \int_{\kappa_{p}^{-}}^{\kappa_{p}^{+}} 
\left|\frac{H_{0}^{(1)}[\kappa(\eta_{0} - \eta_{p})]}{H_{0}^{(1)}[\kappa (\eta_{s} - \eta_{p})]}\right|^{2}
 \kappa^{-7/3} \frac{d\kappa}{S_{h}(\kappa)}\,.
 \nonumber \\
\end{eqnarray}
Notice that the argument of the Hankel function function in the numerator is the only quantity that formally diverges at the pole, so
let us now rewrite the integral as
\begin{eqnarray}
\rho^{2}_{\RL,p} &\simeq& \bar{\cal{E}} 
\int_{\kappa_{p}^{-}}^{\kappa_{p}^{+}} \left|H_{0}^{(1)}[\kappa_{p}(\eta_{0} - \eta_{p})]\right|^{2} d\kappa\,,
\nonumber \\
\bar{\cal{E}} &\equiv& 4 {\cal{D}}^{2} \left|H_{0}^{(1)}[\kappa_{p} (\eta_{s} - \eta_{p})]\right|^{-2} \frac{\kappa_{p}^{-7/3}}{S_{h}(\kappa_{p})}\,,\quad 
\end{eqnarray}
where we have pulled out several factors that are non-singular at the pole and roughly constant near the resonance. 
These expressions lead to singular results if one evaluates it with the late-time expansion of the Hankel function of Eq.~\eqref{prop-exp}. 
Instead,  using the small-argument Hankel function expansion provided in Eq.~\eqref{Hankel-asy}, 
 the SNR integral becomes
\begin{eqnarray}
\rho^{2}_{\RL,p} &\simeq& \bar{\cal{E}} 
\int_{\kappa_{p}^{-}}^{\kappa_{p}^{+}} 
\left\{ {\rm{sign}}[\kappa_{p} {\cal{R}} (\kappa - \kappa_{p})]
  \right.
    \\ 
 & -& \left. \frac{4}{\pi^2} \left[\ln\left(\frac{|\kappa_{p}{\cal{R}} (\kappa - \kappa_{p})|}{2}\right) + \gamma_{\rm
       E}\right]^{2}\right\} d\kappa\,, \nonumber 
\end{eqnarray}
whose most divergent contribution is asymptotic to 
\begin{equation}
\rho^{2}_{\RL,p} \sim  \bar{\cal{E}} \; \epsilon \; \left[\ln\left(\kappa_{p} {\cal{R}} \epsilon \right)\right]^{2}
= \bar{\cal{E}} \; \epsilon \; \left[\ln\left(\frac{\kappa_{p} \epsilon}{\delta \kappa_{p}^{2}} \right)\right]^{2}\,.
\end{equation}
Notice again that the contribution to the SNR near the pole is finite and in fact it formally vanishes in the limit $\delta \kappa_{p} \to 0$. 
As is clear from the above analysis, regularization is here justified: for a suitably small regularization interval
$\epsilon$, the contribution from a neighborhood $\epsilon$ around the pole is small. To provide a numerical
example, saturating the Solar System constraint again ($\kappa_{p} = 150 \; {\rm{Hz}}$) and choosing a regularization interval of size
$\epsilon = 10^{-8} {\rm{Hz}}$ we find that $\rho_{\RL,p} \lesssim 30 \hat{\rho}_{\RL,p}$, where $\hat{\rho}_{\RL,p}$ is the contribution
of the SNR in an interval of size $\delta \kappa_{p}$ in the absence of a resonance. The SNR is mostly dominated by contributions between
$100$-$300 \; {\rm{Hz}}$, so splitting this region into segments of size $\delta f_{p} = 10^{-8} \; {\rm{Hz}}$, one obtains approximately $10^{10}$ segments. 
Assuming each segments carries approximately the same power, one can estimate that $\hat{\rho}_{\RL,p} \sim 10/10^{10} = {\cal{O}}(10^{-9})$. 
Therefore, the contribution from the resonance that we lose by regularizing is approximately $\rho_{\RL,p} \lesssim 10^{-8}$ and can be 
neglected.

\bibliography{pvgw}

\begin{thebibliography}{67}
\expandafter\ifx\csname natexlab\endcsname\relax\def\natexlab#1{#1}\fi
\expandafter\ifx\csname bibnamefont\endcsname\relax
  \def\bibnamefont#1{#1}\fi
\expandafter\ifx\csname bibfnamefont\endcsname\relax
  \def\bibfnamefont#1{#1}\fi
\expandafter\ifx\csname citenamefont\endcsname\relax
  \def\citenamefont#1{#1}\fi
\expandafter\ifx\csname url\endcsname\relax
  \def\url#1{\texttt{#1}}\fi
\expandafter\ifx\csname urlprefix\endcsname\relax\def\urlprefix{URL }\fi
\providecommand{\bibinfo}[2]{#2}
\providecommand{\eprint}[2][]{\url{#2}}

\bibitem[{\citenamefont{Bell and Jackiw}(1969)}]{Bell:1969ts}
\bibinfo{author}{\bibfnamefont{J.~S.} \bibnamefont{Bell}} \bibnamefont{and}
  \bibinfo{author}{\bibfnamefont{R.}~\bibnamefont{Jackiw}},
  \bibinfo{journal}{Nuovo Cim.} \textbf{\bibinfo{volume}{A60}},
  \bibinfo{pages}{47} (\bibinfo{year}{1969}).

\bibitem[{\citenamefont{{Adler}}(1969)}]{1969PhRv..177.2426A}
\bibinfo{author}{\bibfnamefont{S.~L.} \bibnamefont{{Adler}}},
  \bibinfo{journal}{Physical Review} \textbf{\bibinfo{volume}{177}},
  \bibinfo{pages}{2426} (\bibinfo{year}{1969}).

\bibitem[{\citenamefont{Alvarez-Gaume and Witten}(1984)}]{AlvarezGaume:1983ig}
\bibinfo{author}{\bibfnamefont{L.}~\bibnamefont{Alvarez-Gaume}}
  \bibnamefont{and} \bibinfo{author}{\bibfnamefont{E.}~\bibnamefont{Witten}},
  \bibinfo{journal}{Nucl. Phys.} \textbf{\bibinfo{volume}{B234}},
  \bibinfo{pages}{269} (\bibinfo{year}{1984}).

\bibitem[{\citenamefont{Green et~al.}(1987)\citenamefont{Green, Schwarz, and
  Witten}}]{Green:1987mn}
\bibinfo{author}{\bibfnamefont{M.~B.} \bibnamefont{Green}},
  \bibinfo{author}{\bibfnamefont{J.~H.} \bibnamefont{Schwarz}},
  \bibnamefont{and} \bibinfo{author}{\bibfnamefont{E.}~\bibnamefont{Witten}},
  \emph{\bibinfo{title}{Superstring Theory. Vol. 2: Loop Amplitides, Anomalies
  and Phenomenology}} (\bibinfo{publisher}{Cambridge University Press
  (Cambridge Monographs On Mathematical Physics)}, \bibinfo{address}{Cambridge,
  Uk}, \bibinfo{year}{1987}).

\bibitem[{\citenamefont{Alexander and Gates}(2006)}]{Alexander:2004xd}
\bibinfo{author}{\bibfnamefont{S.~H.~S.} \bibnamefont{Alexander}}
  \bibnamefont{and} \bibinfo{author}{\bibfnamefont{J.}~\bibnamefont{Gates},
  \bibfnamefont{S.~James}}, \bibinfo{journal}{JCAP}
  \textbf{\bibinfo{volume}{0606}}, \bibinfo{pages}{018} (\bibinfo{year}{2006}),
  \eprint{hep-th/0409014}.

\bibitem[{\citenamefont{Ashtekar et~al.}(1989)\citenamefont{Ashtekar,
  Balachandran, and Jo}}]{Ashtekar:1988sw}
\bibinfo{author}{\bibfnamefont{A.}~\bibnamefont{Ashtekar}},
  \bibinfo{author}{\bibfnamefont{A.~P.} \bibnamefont{Balachandran}},
  \bibnamefont{and} \bibinfo{author}{\bibfnamefont{S.}~\bibnamefont{Jo}},
  \bibinfo{journal}{Int. J. Mod. Phys. A} \textbf{\bibinfo{volume}{4}},
  \bibinfo{pages}{1493} (\bibinfo{year}{1989}).

\bibitem[{\citenamefont{Taveras and Yunes}(2008)}]{Taveras:2008yf}
\bibinfo{author}{\bibfnamefont{V.}~\bibnamefont{Taveras}} \bibnamefont{and}
  \bibinfo{author}{\bibfnamefont{N.}~\bibnamefont{Yunes}},
  \bibinfo{journal}{Phys. Rev.} \textbf{\bibinfo{volume}{D78}},
  \bibinfo{pages}{064070} (\bibinfo{year}{2008}), \eprint{0807.2652}.

\bibitem[{\citenamefont{Calcagni and Mercuri}(2009)}]{Calcagni:2009xz}
\bibinfo{author}{\bibfnamefont{G.}~\bibnamefont{Calcagni}} \bibnamefont{and}
  \bibinfo{author}{\bibfnamefont{S.}~\bibnamefont{Mercuri}},
  \bibinfo{journal}{Phys. Rev.} \textbf{\bibinfo{volume}{D79}},
  \bibinfo{pages}{084004} (\bibinfo{year}{2009}), \eprint{0902.0957}.

\bibitem[{\citenamefont{Mercuri and Taveras}(2009)}]{Mercuri:2009zt}
\bibinfo{author}{\bibfnamefont{S.}~\bibnamefont{Mercuri}} \bibnamefont{and}
  \bibinfo{author}{\bibfnamefont{V.}~\bibnamefont{Taveras}},
  \bibinfo{journal}{Phys. Rev.} \textbf{\bibinfo{volume}{D80}},
  \bibinfo{pages}{104007} (\bibinfo{year}{2009}), \eprint{0903.4407}.

\bibitem[{\citenamefont{Gates et~al.}(2009)\citenamefont{Gates, Ketov, and
  Yunes}}]{Gates:2009pt}
\bibinfo{author}{\bibfnamefont{S.~J.} \bibnamefont{Gates}, \bibfnamefont{Jr.}},
  \bibinfo{author}{\bibfnamefont{S.~V.} \bibnamefont{Ketov}}, \bibnamefont{and}
  \bibinfo{author}{\bibfnamefont{N.}~\bibnamefont{Yunes}},
  \bibinfo{journal}{Phys. Rev.} \textbf{\bibinfo{volume}{D80}},
  \bibinfo{pages}{065003} (\bibinfo{year}{2009}), \eprint{0906.4978}.

\bibitem[{\citenamefont{Alexander and Yunes}(2009)}]{Alexander:2009tp}
\bibinfo{author}{\bibfnamefont{S.}~\bibnamefont{Alexander}} \bibnamefont{and}
  \bibinfo{author}{\bibfnamefont{N.}~\bibnamefont{Yunes}},
  \bibinfo{journal}{Phys. Rept.} \textbf{\bibinfo{volume}{480}},
  \bibinfo{pages}{1} (\bibinfo{year}{2009}), \eprint{0907.2562}.

\bibitem[{\citenamefont{Jackiw and Pi}(2003)}]{Jackiw:2003pm}
\bibinfo{author}{\bibfnamefont{R.}~\bibnamefont{Jackiw}} \bibnamefont{and}
  \bibinfo{author}{\bibfnamefont{S.~Y.} \bibnamefont{Pi}},
  \bibinfo{journal}{Phys. Rev. D} \textbf{\bibinfo{volume}{68}},
  \bibinfo{pages}{104012} (\bibinfo{year}{2003}).

\bibitem[{\citenamefont{Weinberg}(1996)}]{Weinberg:1996kr}
\bibinfo{author}{\bibfnamefont{S.}~\bibnamefont{Weinberg}},
  \emph{\bibinfo{title}{{The quantum theory of fields, Vol. 2: Modern
  applications}}} (\bibinfo{publisher}{Univ. Pr.}, \bibinfo{address}{Cambridge,
  UK}, \bibinfo{year}{1996}).

\bibitem[{\citenamefont{Polchinski}(1998)}]{Polchinski:1998rr}
\bibinfo{author}{\bibfnamefont{J.}~\bibnamefont{Polchinski}},
  \emph{\bibinfo{title}{{String theory. Vol. 2 Superstring theory and beyond}}}
  (\bibinfo{publisher}{Univ. Pr.}, \bibinfo{address}{Cambridge, UK},
  \bibinfo{year}{1998}).

\bibitem[{\citenamefont{Weinberg}(2008)}]{Weinberg:2008hq}
\bibinfo{author}{\bibfnamefont{S.}~\bibnamefont{Weinberg}},
  \bibinfo{journal}{Phys. Rev.} \textbf{\bibinfo{volume}{D77}},
  \bibinfo{pages}{123541} (\bibinfo{year}{2008}), \eprint{0804.4291}.

\bibitem[{\citenamefont{Contaldi et~al.}(2008)\citenamefont{Contaldi, Magueijo,
  and Smolin}}]{Contaldi:2008yz}
\bibinfo{author}{\bibfnamefont{C.~R.} \bibnamefont{Contaldi}},
  \bibinfo{author}{\bibfnamefont{J.}~\bibnamefont{Magueijo}}, \bibnamefont{and}
  \bibinfo{author}{\bibfnamefont{L.}~\bibnamefont{Smolin}},
  \bibinfo{journal}{Phys. Rev. Lett.} \textbf{\bibinfo{volume}{101}},
  \bibinfo{pages}{141101} (\bibinfo{year}{2008}), \eprint{0806.3082}.

\bibitem[{\citenamefont{Lue et~al.}(1999)\citenamefont{Lue, Wang, and
  Kamionkowski}}]{Lue:1998mq}
\bibinfo{author}{\bibfnamefont{A.}~\bibnamefont{Lue}},
  \bibinfo{author}{\bibfnamefont{L.-M.} \bibnamefont{Wang}}, \bibnamefont{and}
  \bibinfo{author}{\bibfnamefont{M.}~\bibnamefont{Kamionkowski}},
  \bibinfo{journal}{Phys. Rev. Lett.} \textbf{\bibinfo{volume}{83}},
  \bibinfo{pages}{1506} (\bibinfo{year}{1999}).

\bibitem[{\citenamefont{Alexander et~al.}(2008)\citenamefont{Alexander, Finn,
  and Yunes}}]{Alexander:2007kv}
\bibinfo{author}{\bibfnamefont{S.}~\bibnamefont{Alexander}},
  \bibinfo{author}{\bibfnamefont{L.~S.} \bibnamefont{Finn}}, \bibnamefont{and}
  \bibinfo{author}{\bibfnamefont{N.}~\bibnamefont{Yunes}},
  \bibinfo{journal}{Phys. Rev.} \textbf{\bibinfo{volume}{D78}},
  \bibinfo{pages}{066005} (\bibinfo{year}{2008}), \eprint{0712.2542}.

\bibitem[{\citenamefont{Yunes and Finn}(2009)}]{Yunes:2008bu}
\bibinfo{author}{\bibfnamefont{N.}~\bibnamefont{Yunes}} \bibnamefont{and}
  \bibinfo{author}{\bibfnamefont{L.~S.} \bibnamefont{Finn}},
  \bibinfo{journal}{J. Phys. Conf. Ser.} \textbf{\bibinfo{volume}{154}},
  \bibinfo{pages}{012041} (\bibinfo{year}{2009}), \eprint{0811.0181}.

\bibitem[{\citenamefont{Alexander et~al.}(2006)\citenamefont{Alexander, Peskin,
  and Sheikh-Jabbari}}]{Alexander:2004us}
\bibinfo{author}{\bibfnamefont{S.~H.-S.} \bibnamefont{Alexander}},
  \bibinfo{author}{\bibfnamefont{M.~E.} \bibnamefont{Peskin}},
  \bibnamefont{and} \bibinfo{author}{\bibfnamefont{M.~M.}
  \bibnamefont{Sheikh-Jabbari}}, \bibinfo{journal}{Phys. Rev. Lett.}
  \textbf{\bibinfo{volume}{96}}, \bibinfo{pages}{081301}
  (\bibinfo{year}{2006}).

\bibitem[{\citenamefont{Alexander and Martin}(2005)}]{Alexander:2004wk}
\bibinfo{author}{\bibfnamefont{S.}~\bibnamefont{Alexander}} \bibnamefont{and}
  \bibinfo{author}{\bibfnamefont{J.}~\bibnamefont{Martin}},
  \bibinfo{journal}{Phys. Rev. D} \textbf{\bibinfo{volume}{71}},
  \bibinfo{pages}{063526} (\bibinfo{year}{2005}).

\bibitem[{\citenamefont{Saito et~al.}(2007)\citenamefont{Saito, Ichiki, and
  Taruya}}]{Saito:2007kt}
\bibinfo{author}{\bibfnamefont{S.}~\bibnamefont{Saito}},
  \bibinfo{author}{\bibfnamefont{K.}~\bibnamefont{Ichiki}}, \bibnamefont{and}
  \bibinfo{author}{\bibfnamefont{A.}~\bibnamefont{Taruya}},
  \bibinfo{journal}{JCAP} \textbf{\bibinfo{volume}{0709}}, \bibinfo{pages}{002}
  (\bibinfo{year}{2007}).

\bibitem[{\citenamefont{Seto and Taruya}(2007)}]{Seto:2007tn}
\bibinfo{author}{\bibfnamefont{N.}~\bibnamefont{Seto}} \bibnamefont{and}
  \bibinfo{author}{\bibfnamefont{A.}~\bibnamefont{Taruya}},
  \bibinfo{journal}{Phys. Rev. Lett.} \textbf{\bibinfo{volume}{99}},
  \bibinfo{pages}{121101} (\bibinfo{year}{2007}).

\bibitem[{\citenamefont{Satoh et~al.}(2008)\citenamefont{Satoh, Kanno, and
  Soda}}]{Satoh:2007gn}
\bibinfo{author}{\bibfnamefont{M.}~\bibnamefont{Satoh}},
  \bibinfo{author}{\bibfnamefont{S.}~\bibnamefont{Kanno}}, \bibnamefont{and}
  \bibinfo{author}{\bibfnamefont{J.}~\bibnamefont{Soda}},
  \bibinfo{journal}{Phys. Rev. D} \textbf{\bibinfo{volume}{77}},
  \bibinfo{pages}{023526} (\bibinfo{year}{2008}).

\bibitem[{\citenamefont{Laguna et~al.}(2010)\citenamefont{Laguna, Larson,
  Spergel, and Yunes}}]{Laguna:2009re}
\bibinfo{author}{\bibfnamefont{P.}~\bibnamefont{Laguna}},
  \bibinfo{author}{\bibfnamefont{S.~L.} \bibnamefont{Larson}},
  \bibinfo{author}{\bibfnamefont{D.}~\bibnamefont{Spergel}}, \bibnamefont{and}
  \bibinfo{author}{\bibfnamefont{N.}~\bibnamefont{Yunes}},
  \bibinfo{journal}{Astrophys. J. Letts.} \textbf{\bibinfo{volume}{715}},
  \bibinfo{pages}{L12} (\bibinfo{year}{2010}), \eprint{0905.1908}.

\bibitem[{\citenamefont{Yunes and Sopuerta}(2008)}]{Yunes:2007ss}
\bibinfo{author}{\bibfnamefont{N.}~\bibnamefont{Yunes}} \bibnamefont{and}
  \bibinfo{author}{\bibfnamefont{C.~F.} \bibnamefont{Sopuerta}},
  \bibinfo{journal}{Phys. Rev.} \textbf{\bibinfo{volume}{D77}},
  \bibinfo{pages}{064007} (\bibinfo{year}{2008}), \eprint{0712.1028}.

\bibitem[{\citenamefont{Alexander and Yunes}(2008)}]{Alexander:2008wi}
\bibinfo{author}{\bibfnamefont{S.}~\bibnamefont{Alexander}} \bibnamefont{and}
  \bibinfo{author}{\bibfnamefont{N.}~\bibnamefont{Yunes}},
  \bibinfo{journal}{Phys. Rev.} \textbf{\bibinfo{volume}{D77}},
  \bibinfo{pages}{124040} (\bibinfo{year}{2008}), \eprint{0804.1797}.

\bibitem[{\citenamefont{Sopuerta and Yunes}(2009)}]{Sopuerta:2009iy}
\bibinfo{author}{\bibfnamefont{C.~F.} \bibnamefont{Sopuerta}} \bibnamefont{and}
  \bibinfo{author}{\bibfnamefont{N.}~\bibnamefont{Yunes}},
  \bibinfo{journal}{Phys. Rev.} \textbf{\bibinfo{volume}{D80}},
  \bibinfo{pages}{064006} (\bibinfo{year}{2009}), \eprint{0904.4501}.

\bibitem[{\citenamefont{Gehrels et~al.}(2004)}]{Gehrels:2004am}
\bibinfo{author}{\bibfnamefont{N.}~\bibnamefont{Gehrels}} \bibnamefont{et~al.},
  \bibinfo{journal}{AIP Conf. Proc.} \textbf{\bibinfo{volume}{727}},
  \bibinfo{pages}{637} (\bibinfo{year}{2004}).

\bibitem[{\citenamefont{Carson}(2007)}]{Carson:2006af}
\bibinfo{author}{\bibfnamefont{J.~E.} \bibnamefont{Carson}},
  \bibinfo{journal}{J. Phys. Conf. Ser.} \textbf{\bibinfo{volume}{60}},
  \bibinfo{pages}{115} (\bibinfo{year}{2007}).

\bibitem[{\citenamefont{Abbott et~al.}(2009)}]{Abbott:2007kv}
\bibinfo{author}{\bibfnamefont{B.}~\bibnamefont{Abbott}} \bibnamefont{et~al.},
  \bibinfo{journal}{Rept. Prog. Phys.} \textbf{\bibinfo{volume}{72}},
  \bibinfo{pages}{076901} (\bibinfo{year}{2009}), \eprint{0711.3041}.

\bibitem[{\citenamefont{Acernese et~al.}(2007)}]{Acernese:2007zze}
\bibinfo{author}{\bibfnamefont{F.}~\bibnamefont{Acernese}} \bibnamefont{et~al.}
  (\bibinfo{collaboration}{Virgo}), \bibinfo{journal}{Class. Quant. Grav.}
  \textbf{\bibinfo{volume}{24}}, \bibinfo{pages}{S381} (\bibinfo{year}{2007}).

\bibitem[{\citenamefont{Seto}(2007)}]{Seto:2005xb}
\bibinfo{author}{\bibfnamefont{N.}~\bibnamefont{Seto}}, \bibinfo{journal}{Phys.
  Rev. D} \textbf{\bibinfo{volume}{75}}, \bibinfo{pages}{024016}
  (\bibinfo{year}{2007}).

\bibitem[{\citenamefont{Dalal et~al.}(2006)\citenamefont{Dalal, Holz, Hughes,
  and Jain}}]{Dalal:2006qt}
\bibinfo{author}{\bibfnamefont{N.}~\bibnamefont{Dalal}},
  \bibinfo{author}{\bibfnamefont{D.~E.} \bibnamefont{Holz}},
  \bibinfo{author}{\bibfnamefont{S.~A.} \bibnamefont{Hughes}},
  \bibnamefont{and} \bibinfo{author}{\bibfnamefont{B.}~\bibnamefont{Jain}},
  \bibinfo{journal}{Phys. Rev. D} \textbf{\bibinfo{volume}{74}},
  \bibinfo{pages}{063006} (\bibinfo{year}{2006}).

\bibitem[{\citenamefont{Nissanke et~al.}(2009)\citenamefont{Nissanke, Hughes,
  Holz, Dalal, and Sievers}}]{Nissanke:2009kt}
\bibinfo{author}{\bibfnamefont{S.}~\bibnamefont{Nissanke}},
  \bibinfo{author}{\bibfnamefont{S.~A.} \bibnamefont{Hughes}},
  \bibinfo{author}{\bibfnamefont{D.~E.} \bibnamefont{Holz}},
  \bibinfo{author}{\bibfnamefont{N.}~\bibnamefont{Dalal}}, \bibnamefont{and}
  \bibinfo{author}{\bibfnamefont{J.~L.} \bibnamefont{Sievers}}
  (\bibinfo{year}{2009}), \eprint{0904.1017}.

\bibitem[{\citenamefont{Alexander and
  Yunes}(2007{\natexlab{a}})}]{Alexander:2007zg}
\bibinfo{author}{\bibfnamefont{S.}~\bibnamefont{Alexander}} \bibnamefont{and}
  \bibinfo{author}{\bibfnamefont{N.}~\bibnamefont{Yunes}},
  \bibinfo{journal}{Phys. Rev. Lett.} \textbf{\bibinfo{volume}{99}},
  \bibinfo{pages}{241101} (\bibinfo{year}{2007}{\natexlab{a}}).

\bibitem[{\citenamefont{Alexander and
  Yunes}(2007{\natexlab{b}})}]{Alexander:2007vt}
\bibinfo{author}{\bibfnamefont{S.}~\bibnamefont{Alexander}} \bibnamefont{and}
  \bibinfo{author}{\bibfnamefont{N.}~\bibnamefont{Yunes}},
  \bibinfo{journal}{Phys. Rev.} \textbf{\bibinfo{volume}{D75}},
  \bibinfo{pages}{124022} (\bibinfo{year}{2007}{\natexlab{b}}).

\bibitem[{\citenamefont{Smith et~al.}(2008)\citenamefont{Smith, Erickcek,
  Caldwell, and Kamionkowski}}]{Smith:2007jm}
\bibinfo{author}{\bibfnamefont{T.~L.} \bibnamefont{Smith}},
  \bibinfo{author}{\bibfnamefont{A.~L.} \bibnamefont{Erickcek}},
  \bibinfo{author}{\bibfnamefont{R.~R.} \bibnamefont{Caldwell}},
  \bibnamefont{and}
  \bibinfo{author}{\bibfnamefont{M.}~\bibnamefont{Kamionkowski}},
  \bibinfo{journal}{Phys. Rev. D} \textbf{\bibinfo{volume}{77}},
  \bibinfo{pages}{024015} (\bibinfo{year}{2008}).

\bibitem[{\citenamefont{Yunes and Spergel}(2009)}]{Yunes:2008ua}
\bibinfo{author}{\bibfnamefont{N.}~\bibnamefont{Yunes}} \bibnamefont{and}
  \bibinfo{author}{\bibfnamefont{D.~N.} \bibnamefont{Spergel}},
  \bibinfo{journal}{Phys. Rev.} \textbf{\bibinfo{volume}{D80}},
  \bibinfo{pages}{042004} (\bibinfo{year}{2009}), \eprint{0810.5541}.

\bibitem[{\citenamefont{Misner et~al.}(1979)\citenamefont{Misner, Thorne, and
  Wheeler}}]{Misner:1974qy}
\bibinfo{author}{\bibfnamefont{C.~W.} \bibnamefont{Misner}},
  \bibinfo{author}{\bibfnamefont{K.~S.} \bibnamefont{Thorne}},
  \bibnamefont{and} \bibinfo{author}{\bibfnamefont{J.~A.}
  \bibnamefont{Wheeler}} (\bibinfo{year}{1979}), \bibinfo{note}{san Francisco
  1973, 1279p}.

\bibitem[{\citenamefont{Gukov et~al.}(2004)\citenamefont{Gukov, Kachru, Liu,
  and McAllister}}]{Gukov:2003cy}
\bibinfo{author}{\bibfnamefont{S.}~\bibnamefont{Gukov}},
  \bibinfo{author}{\bibfnamefont{S.}~\bibnamefont{Kachru}},
  \bibinfo{author}{\bibfnamefont{X.}~\bibnamefont{Liu}}, \bibnamefont{and}
  \bibinfo{author}{\bibfnamefont{L.}~\bibnamefont{McAllister}},
  \bibinfo{journal}{Phys. Rev. D} \textbf{\bibinfo{volume}{69}},
  \bibinfo{pages}{086008} (\bibinfo{year}{2004}).

\bibitem[{\citenamefont{Abadie et~al.}(2010)}]{Abadie:2010cf}
\bibinfo{author}{\bibfnamefont{J.}~\bibnamefont{Abadie}} \bibnamefont{et~al.}
  (\bibinfo{year}{2010}), \eprint{1003.2480}.

\bibitem[{\citenamefont{{LIGO Scientific
  Collaboration}}(2009)}]{AdvancedLIGONoise}
\bibinfo{author}{\bibnamefont{{LIGO Scientific Collaboration}}},
  \bibinfo{type}{Tech. Rep.}, \bibinfo{institution}{LIGO-T0900288-v3}
  (\bibinfo{year}{2009}),
  \urlprefix\url{https://dcc.ligo.org/cgi-bin/DocDB/ShowDocument?docid=2974}.

\bibitem[{\citenamefont{{Harry} and {the LIGO Scientific
  Collaboration}}(2010)}]{2010CQGra..27h4006H}
\bibinfo{author}{\bibfnamefont{G.~M.} \bibnamefont{{Harry}}} \bibnamefont{and}
  \bibinfo{author}{\bibnamefont{{the LIGO Scientific Collaboration}}},
  \bibinfo{journal}{Classical and Quantum Gravity}
  \textbf{\bibinfo{volume}{27}}, \bibinfo{pages}{084006}
  (\bibinfo{year}{2010}).

\bibitem[{\citenamefont{Yunes and Pretorius}(2009)}]{Yunes:2009ke}
\bibinfo{author}{\bibfnamefont{N.}~\bibnamefont{Yunes}} \bibnamefont{and}
  \bibinfo{author}{\bibfnamefont{F.}~\bibnamefont{Pretorius}},
  \bibinfo{journal}{Phys. Rev.} \textbf{\bibinfo{volume}{D80}},
  \bibinfo{pages}{122003} (\bibinfo{year}{2009}), \eprint{0909.3328}.

\bibitem[{\citenamefont{{Cutler} and {Flanagan}}(1994)}]{CutlerFlanagan:1994}
\bibinfo{author}{\bibfnamefont{C.}~\bibnamefont{{Cutler}}} \bibnamefont{and}
  \bibinfo{author}{\bibfnamefont{E.}~\bibnamefont{{Flanagan}}},
  \bibinfo{journal}{\prd} \textbf{\bibinfo{volume}{49}}, \bibinfo{pages}{2658}
  (\bibinfo{year}{1994}), \eprint{gr-qc/9402014}.

\bibitem[{\citenamefont{Abbott et~al.}(2008{\natexlab{a}})}]{Abbott:2007xi}
\bibinfo{author}{\bibfnamefont{B.}~\bibnamefont{Abbott}} \bibnamefont{et~al.},
  \bibinfo{journal}{Phys. Rev.} \textbf{\bibinfo{volume}{D77}},
  \bibinfo{pages}{062002} (\bibinfo{year}{2008}{\natexlab{a}}),
  \eprint{0704.3368}.

\bibitem[{\citenamefont{Stuver and Finn}(2006)}]{Stuver:2006gr}
\bibinfo{author}{\bibfnamefont{A.~L.} \bibnamefont{Stuver}} \bibnamefont{and}
  \bibinfo{author}{\bibfnamefont{L.~S.} \bibnamefont{Finn}},
  \bibinfo{journal}{Class. Quant. Grav.} \textbf{\bibinfo{volume}{23}},
  \bibinfo{pages}{S799} (\bibinfo{year}{2006}).

\bibitem[{\citenamefont{{Grandcl{\'e}ment}
  et~al.}(2004)\citenamefont{{Grandcl{\'e}ment}, {Ihm}, {Kalogera}, and
  {Belczynski}}}]{2004PhRvD..69j2002G}
\bibinfo{author}{\bibfnamefont{P.}~\bibnamefont{{Grandcl{\'e}ment}}},
  \bibinfo{author}{\bibfnamefont{M.}~\bibnamefont{{Ihm}}},
  \bibinfo{author}{\bibfnamefont{V.}~\bibnamefont{{Kalogera}}},
  \bibnamefont{and}
  \bibinfo{author}{\bibfnamefont{K.}~\bibnamefont{{Belczynski}}},
  \bibinfo{journal}{\prd} \textbf{\bibinfo{volume}{69}},
  \bibinfo{pages}{102002} (\bibinfo{year}{2004}), \eprint{arXiv:gr-qc/0312084}.

\bibitem[{\citenamefont{{Soderberg} et~al.}(2006)\citenamefont{{Soderberg},
  {Berger}, {Kasliwal}, {Frail}, {Price}, {Schmidt}, {Kulkarni}, {Fox},
  {Cenko}, {Gal-Yam} et~al.}}]{2006ApJ...650..261S}
\bibinfo{author}{\bibfnamefont{A.~M.} \bibnamefont{{Soderberg}}},
  \bibinfo{author}{\bibfnamefont{E.}~\bibnamefont{{Berger}}},
  \bibinfo{author}{\bibfnamefont{M.}~\bibnamefont{{Kasliwal}}},
  \bibinfo{author}{\bibfnamefont{D.~A.} \bibnamefont{{Frail}}},
  \bibinfo{author}{\bibfnamefont{P.~A.} \bibnamefont{{Price}}},
  \bibinfo{author}{\bibfnamefont{B.~P.} \bibnamefont{{Schmidt}}},
  \bibinfo{author}{\bibfnamefont{S.~R.} \bibnamefont{{Kulkarni}}},
  \bibinfo{author}{\bibfnamefont{D.~B.} \bibnamefont{{Fox}}},
  \bibinfo{author}{\bibfnamefont{S.~B.} \bibnamefont{{Cenko}}},
  \bibinfo{author}{\bibfnamefont{A.}~\bibnamefont{{Gal-Yam}}},
  \bibnamefont{et~al.}, \bibinfo{journal}{\apj} \textbf{\bibinfo{volume}{650}},
  \bibinfo{pages}{261} (\bibinfo{year}{2006}), \eprint{astro-ph/0601455}.

\bibitem[{\citenamefont{{Grupe} et~al.}(2006)\citenamefont{{Grupe}, {Burrows},
  {Patel}, {Kouveliotou}, {Zhang}, {M{\'e}sz{\'a}ros}, {Wijers}, and
  {Gehrels}}}]{2006ApJ...653..462G}
\bibinfo{author}{\bibfnamefont{D.}~\bibnamefont{{Grupe}}},
  \bibinfo{author}{\bibfnamefont{D.~N.} \bibnamefont{{Burrows}}},
  \bibinfo{author}{\bibfnamefont{S.~K.} \bibnamefont{{Patel}}},
  \bibinfo{author}{\bibfnamefont{C.}~\bibnamefont{{Kouveliotou}}},
  \bibinfo{author}{\bibfnamefont{B.}~\bibnamefont{{Zhang}}},
  \bibinfo{author}{\bibfnamefont{P.}~\bibnamefont{{M{\'e}sz{\'a}ros}}},
  \bibinfo{author}{\bibfnamefont{R.~A.~M.} \bibnamefont{{Wijers}}},
  \bibnamefont{and}
  \bibinfo{author}{\bibfnamefont{N.}~\bibnamefont{{Gehrels}}},
  \bibinfo{journal}{\apj} \textbf{\bibinfo{volume}{653}}, \bibinfo{pages}{462}
  (\bibinfo{year}{2006}), \eprint{astro-ph/0603773}.

\bibitem[{\citenamefont{Van Den~Broeck and
  Sengupta}(2007)}]{VanDenBroeck:2006qu}
\bibinfo{author}{\bibfnamefont{C.}~\bibnamefont{Van Den~Broeck}}
  \bibnamefont{and} \bibinfo{author}{\bibfnamefont{A.~S.}
  \bibnamefont{Sengupta}}, \bibinfo{journal}{Class. Quant. Grav.}
  \textbf{\bibinfo{volume}{24}}, \bibinfo{pages}{155} (\bibinfo{year}{2007}),
  \eprint{gr-qc/0607092}.

\bibitem[{\citenamefont{Jarosik et~al.}(2010)}]{Jarosik:2010iu}
\bibinfo{author}{\bibfnamefont{N.}~\bibnamefont{Jarosik}} \bibnamefont{et~al.}
  (\bibinfo{year}{2010}), \eprint{1001.4744}.

\bibitem[{\citenamefont{{Menou} et~al.}(2008)\citenamefont{{Menou}, {Haiman},
  and {Kocsis}}}]{2008NewAR..51..884M}
\bibinfo{author}{\bibfnamefont{K.}~\bibnamefont{{Menou}}},
  \bibinfo{author}{\bibfnamefont{Z.}~\bibnamefont{{Haiman}}}, \bibnamefont{and}
  \bibinfo{author}{\bibfnamefont{B.}~\bibnamefont{{Kocsis}}},
  \bibinfo{journal}{New Astronomy Review} \textbf{\bibinfo{volume}{51}},
  \bibinfo{pages}{884} (\bibinfo{year}{2008}), \eprint{0803.3627}.

\bibitem[{\citenamefont{Gal-Yam et~al.}(2008)}]{GalYam:2005jf}
\bibinfo{author}{\bibfnamefont{A.}~\bibnamefont{Gal-Yam}} \bibnamefont{et~al.},
  \bibinfo{journal}{Astrophys. J.} \textbf{\bibinfo{volume}{686}},
  \bibinfo{pages}{408} (\bibinfo{year}{2008}), \eprint{astro-ph/0509891}.

\bibitem[{\citenamefont{{Nakar}}(2007)}]{NakarReviewArticle2006}
\bibinfo{author}{\bibfnamefont{E.}~\bibnamefont{{Nakar}}},
  \bibinfo{journal}{Phys. Rept.} \textbf{\bibinfo{volume}{442}},
  \bibinfo{pages}{166} (\bibinfo{year}{2007}), \eprint{arXiv:astro-ph/0701748}.

\bibitem[{\citenamefont{{Gehrels} et~al.}(2007)\citenamefont{{Gehrels},
  {Cannizzo}, and {Norris}}}]{Gehrels-shortgrb-SwiftReview-Mid2007}
\bibinfo{author}{\bibfnamefont{N.}~\bibnamefont{{Gehrels}}},
  \bibinfo{author}{\bibfnamefont{J.~K.} \bibnamefont{{Cannizzo}}},
  \bibnamefont{and} \bibinfo{author}{\bibfnamefont{J.~P.}
  \bibnamefont{{Norris}}}, \bibinfo{journal}{New Journal of Physics}
  \textbf{\bibinfo{volume}{9}}, \bibinfo{pages}{37} (\bibinfo{year}{2007}).

\bibitem[{\citenamefont{{Zemp} et~al.}(2009)\citenamefont{{Zemp},
  {Ramirez-Ruiz}, and {Diemand}}}]{2009ApJ...705L.186Z}
\bibinfo{author}{\bibfnamefont{M.}~\bibnamefont{{Zemp}}},
  \bibinfo{author}{\bibfnamefont{E.}~\bibnamefont{{Ramirez-Ruiz}}},
  \bibnamefont{and}
  \bibinfo{author}{\bibfnamefont{J.}~\bibnamefont{{Diemand}}},
  \bibinfo{journal}{\apjl} \textbf{\bibinfo{volume}{705}},
  \bibinfo{pages}{L186} (\bibinfo{year}{2009}), \eprint{0910.1593}.

\bibitem[{\citenamefont{{O'Shaughnessy}
  et~al.}(2008)\citenamefont{{O'Shaughnessy}, {Kalogera}, and
  {Belczynski}}}]{PSgrbs-popsyn}
\bibinfo{author}{\bibfnamefont{R.}~\bibnamefont{{O'Shaughnessy}}},
  \bibinfo{author}{\bibfnamefont{V.}~\bibnamefont{{Kalogera}}},
  \bibnamefont{and}
  \bibinfo{author}{\bibfnamefont{C.}~\bibnamefont{{Belczynski}}},
  \bibinfo{journal}{ApJ} \textbf{\bibinfo{volume}{675}}, \bibinfo{pages}{566+}
  (\bibinfo{year}{2008}).

\bibitem[{\citenamefont{{Paciesas} et~al.}(1999)\citenamefont{{Paciesas},
  {Meegan}, {Pendleton}, {Briggs}, {Kouveliotou}, {Koshut}, {Lestrade},
  {McCollough}, {Brainerd}, {Hakkila} et~al.}}]{1999ApJS..122..465P}
\bibinfo{author}{\bibfnamefont{W.~S.} \bibnamefont{{Paciesas}}},
  \bibinfo{author}{\bibfnamefont{C.~A.} \bibnamefont{{Meegan}}},
  \bibinfo{author}{\bibfnamefont{G.~N.} \bibnamefont{{Pendleton}}},
  \bibinfo{author}{\bibfnamefont{M.~S.} \bibnamefont{{Briggs}}},
  \bibinfo{author}{\bibfnamefont{C.}~\bibnamefont{{Kouveliotou}}},
  \bibinfo{author}{\bibfnamefont{T.~M.} \bibnamefont{{Koshut}}},
  \bibinfo{author}{\bibfnamefont{J.~P.} \bibnamefont{{Lestrade}}},
  \bibinfo{author}{\bibfnamefont{M.~L.} \bibnamefont{{McCollough}}},
  \bibinfo{author}{\bibfnamefont{J.~J.} \bibnamefont{{Brainerd}}},
  \bibinfo{author}{\bibfnamefont{J.}~\bibnamefont{{Hakkila}}},
  \bibnamefont{et~al.}, \bibinfo{journal}{Astrophys. J. Suppl.}
  \textbf{\bibinfo{volume}{122}}, \bibinfo{pages}{465} (\bibinfo{year}{1999}),
  \eprint{arXiv:astro-ph/9903205}.

\bibitem[{\citenamefont{{Guetta} and {Piran}}(2005)}]{GuettaPiran2005}
\bibinfo{author}{\bibfnamefont{D.}~\bibnamefont{{Guetta}}} \bibnamefont{and}
  \bibinfo{author}{\bibfnamefont{T.}~\bibnamefont{{Piran}}},
  \bibinfo{journal}{Astron. Astrophys.} \textbf{\bibinfo{volume}{435}},
  \bibinfo{pages}{421} (\bibinfo{year}{2005}).

\bibitem[{\citenamefont{{Guetta} and {Piran}}(2006)}]{GuettaPiran2006}
\bibinfo{author}{\bibfnamefont{D.}~\bibnamefont{{Guetta}}} \bibnamefont{and}
  \bibinfo{author}{\bibfnamefont{T.}~\bibnamefont{{Piran}}},
  \bibinfo{journal}{Astron. Astrophys.} \textbf{\bibinfo{volume}{453}},
  \bibinfo{pages}{823} (\bibinfo{year}{2006}).

\bibitem[{\citenamefont{Abbott et~al.}(2005)}]{Abbott:2005yy}
\bibinfo{author}{\bibfnamefont{B.}~\bibnamefont{Abbott}} \bibnamefont{et~al.},
  \bibinfo{journal}{Phys. Rev.} \textbf{\bibinfo{volume}{D72}},
  \bibinfo{pages}{042002} (\bibinfo{year}{2005}), \eprint{gr-qc/0501068}.

\bibitem[{\citenamefont{Abbott et~al.}(2008{\natexlab{b}})}]{Abbott:2008zzb}
\bibinfo{author}{\bibfnamefont{B.}~\bibnamefont{Abbott}} \bibnamefont{et~al.},
  \bibinfo{journal}{Phys. Rev.} \textbf{\bibinfo{volume}{D77}},
  \bibinfo{pages}{062004} (\bibinfo{year}{2008}{\natexlab{b}}),
  \eprint{0709.0766}.

\bibitem[{\citenamefont{O'Shaughnessy et~al.}(2008)\citenamefont{O'Shaughnessy,
  Kim, Kalogera, and Belczynski}}]{O'Shaughnessy:2006wh}
\bibinfo{author}{\bibfnamefont{R.}~\bibnamefont{O'Shaughnessy}},
  \bibinfo{author}{\bibfnamefont{C.}~\bibnamefont{Kim}},
  \bibinfo{author}{\bibfnamefont{V.}~\bibnamefont{Kalogera}}, \bibnamefont{and}
  \bibinfo{author}{\bibfnamefont{K.}~\bibnamefont{Belczynski}},
  \bibinfo{journal}{Astrophys. J.} \textbf{\bibinfo{volume}{672}},
  \bibinfo{pages}{479} (\bibinfo{year}{2008}).

\bibitem[{\citenamefont{{O'Shaughnessy}
  et~al.}(2009)\citenamefont{{O'Shaughnessy}, {Kalogera}, and
  {Belcynski}}}]{PSellipticals}
\bibinfo{author}{\bibfnamefont{R.}~\bibnamefont{{O'Shaughnessy}}},
  \bibinfo{author}{\bibfnamefont{V.}~\bibnamefont{{Kalogera}}},
  \bibnamefont{and}
  \bibinfo{author}{\bibfnamefont{K.}~\bibnamefont{{Belcynski}}},
  \bibinfo{journal}{in preparation}  (\bibinfo{year}{2009}).

\bibitem[{\citenamefont{Bender and Orszag}(1999)}]{Bender}
\bibinfo{author}{\bibfnamefont{C.~M.} \bibnamefont{Bender}} \bibnamefont{and}
  \bibinfo{author}{\bibfnamefont{S.~A.} \bibnamefont{Orszag}},
  \emph{\bibinfo{title}{Advanced mathematical methods for scientists and
  engineers 1, Asymptotic methods and perturbation theory}}
  (\bibinfo{publisher}{Springer}, \bibinfo{address}{New York},
  \bibinfo{year}{1999}).

\end{thebibliography}
\end{document}